\documentclass[fleqn]{iac}

\usepackage[table]{xcolor}
\usepackage[nolist, nohyperlinks]{acronym}
\usepackage[hyphens]{url}
\usepackage[hidelinks]{hyperref} 
\usepackage{multirow}
\hypersetup{breaklinks=true}
\usepackage{longtable,tabularx}
\usepackage{amsmath}
\usepackage{physics}

\usepackage{caption}
\usepackage{subcaption}
\usepackage{enumitem}

\makeatletter
\def\tagform@#1{\maketag@@@{(\ignorespaces#1\unskip\@@italiccorr)}}
\makeatother

\usepackage{makecell}

\definecolor{lightbg}{rgb}{0.9, 0.9, 0.9}
\definecolor{lightgray}{rgb}{0.95, 0.95, 0.95}

\usepackage{hyperref}

\begin{document}

\IACpaperyear{22}
\IACpapernumber{B2.2.3}

\title{Dynamic Frequency Assignment for Mobile Users in Multibeam Satellite Constellations}

\IACauthor{{Guillem Casadesus-Vila$^a$, Juan Jose Garau-Luis$^a$, Nils Pachler$^a$, Edward Crawley$^a$, Bruce Cameron$^a$}}{$^a$ Department of Aeronautics and Astronautics, Massachusetts Institute of Technology, 77 Massachusetts Avenue, Cambridge 02139, United States, \{guillemc, garau, pachler, crawley, bcameron\}@mit.edu}

\abstract{

    Mobile users such as airplanes or ships will constitute an important segment of the future satellite communications market. Operators are now able to leverage digital payloads that allow flexible resource allocation policies that are robust against dynamic user bases. One of the key problems is managing the frequency spectrum efficiently, which has not been sufficiently explored for mobile users.
    To address this gap, we propose a dynamic frequency management algorithm based on linear programming that assigns resources in scenarios with both fixed and mobile users by combining long-term planning with real-time operation. We propose different strategies divided into proactive strategies, which stem from robust optimization practices, and reactive strategies, which exploit a high degree of real-time control. This represents a tradeoff between how conservative long-time planning should be and how much real-time reconfiguration is needed.
    To assess the performance of our method and to determine which proactive and reactive strategies work better under which context, we simulate operational use cases of non-geostationary constellations with different levels of dimensionality and uncertainty, showing that our method is able to serve over 99.97\% of the fixed and mobile users in scenarios with more than 900 beams. Finally, we discuss the trade-offs between the studied strategies, in terms of number of served users, power consumption, and number of changes that need to happen during operations.
}

\maketitle


\section{Introduction}

\subsection{Motivation}



Satellite communication has become a promising solution to satisfy society's growing need to be connected anytime, anywhere, and even on the move \cite{NorthernSkyResearch2022, NorthernSkyResearch2022b}. To meet the market necessities, both established satellite operators, such as SES S.A., which currently operates a constellation in medium Earth orbit (MEO), as well as newer contestants, such as SpaceX or Amazon, are developing the next generation of non-geostationary orbit (NGSO) constellations \cite{Pachler2021a}.




Mobile users (including aeronautical, maritime, and land vehicles) transit areas where connectivity using ground infrastructure might be unreliable or completely unavailable, and therefore other networks such as satellite communications come into play. Due to the increase in the number of users, this segment is expected to represent 40\% of the cumulative market revenue throughout the next decade \cite{NorthernSkyResearch2022, NorthernSkyResearch2022b, NorthernSkyResearch2022c}.
In contrast to past satellite communications with mobile platforms, which supported only low data rate services, mobile users currently require data rates comparable to that of fixed terminals \cite{InternationalTelecommunicationUnion2022}.




To meet the increased broadband connectivity needs, operators will rely on the flexibility of modern payloads. Many new satellites are equipped with thousands of spot beams that can be steered to cover selected service areas and track users, and whose frequency, bandwidth, and power can be adapted in real-time according to the users' demands \cite{SESS.A., SpaceExplorationHoldingsLLC}. This technology enables efficient spectrum usage by reusing frequency through spatial separation between beams and employing different polarizations. Additionally, it allows for reducing the number of handovers,
compared to grid layouts or non-steerable beams, which further improves constellation efficiency \cite{Al-Hraishawi2021, Markovitz2022}.

This flexibility comes at the cost of additional decision-making. The management of constellation resources can be grouped into four tasks: deciding how many beams to use and where to place them, routing each beam to a gateway through a satellite, assigning a certain amount of bandwidth within the available frequency spectrum, and, finally, powering each beam \cite{Garau-Luis2021}. Finding a feasible distribution of resources that satisfies the changing demand needs of a user base is known as the Dynamic Resource Management (DRM) problem and is well-known in the community \cite{Garau-Luis2021,Guerster2019}.



The demand and operational characteristics of mobile users increase the complexity of the DRM problem. Firstly, they contribute to a geographical and temporal unbalance of traffic demand: it is estimated that 50\% of the aeronautical traffic is concentrated in 4\% of Earth's surface, and 80\% of the maritime traffic is concentrated in 15\% of it \cite{Panthi2016}. Entire aircraft fleets can fly from one region to another and back multiple times in the same day, contributing to large demand peaks \cite{McLain2013}. If resource management approaches do not consider route-specific traffic, supporting aeronautical and maritime services will be inefficient and costly, achieving constellation efficiencies below 5\% \cite{McLain2017}, which stem from the necessity to track the users while avoiding interference with the ground infrastructure.

Secondly, mobile users are an understudied source of uncertainty, with demand requests at unanticipated times and locations that have substantial implications for interference control, an essential component of spectrum management. For example, aeronautical users can suffer from delayed flight departures or trajectory changes, resulting in harmful interference with other users or undesired network saturation. Other users, such as trucks, might require service in a specific location without previous notice, further contributing to uncertain and time-varying demand distributions.



The demand dynamics, uncertainty, and operational challenges introduced by mobile users make the allocation of frequency resources an especially challenging problem, which only worsens when constellations scale to thousands of satellites and beams \cite{Pachler2021a}. To enable efficient resource utilization in the upcoming satellite communications landscape, future frequency assignment techniques need to account for the additional layers of complexity introduced by mobile users.

\subsection{Literature review}


Due to the renewed interest in providing broadband connectivity from space, the DRM problem has received the attention of many researchers over the last few years. From a system's perspective, the authors in \cite{Guerster2019} present a problem representation and highlight the importance of two functional blocks: an offline planner without computational time constraints that can exhaustively explore the solution space and a real-time optimizer that modifies the resource allocation based on timely information. The authors stress the need to generate a baseline plan with information known before operations, which includes a certain conservatism to account for the uncertainty present in the planning and enable the real-time optimizer to reallocate resources successfully. In \cite{Garau-Luis2021}, the authors reinforce the idea of leveraging the flexibility of satellite payloads with this two-step allocation and discuss its implications.


Within the DRM problem's technical approach, frequency assignment is one of the most studied tasks. Early works focused on minimizing co-channel interference by rearranging the allocation of a set of carriers allocated to another system, considering fixed terminal positions and GEO satellites. The segmentation of the available spectrum was proposed in \cite{Mizuike1989}, proving that the problem is NP-complete when the allocations need to occupy multiple adjacent segments. In the same context, later works proposed neural networks \cite{Funabiki1997, Wang2011} and the use of other metaheuristics such as evolutionary algorithms \cite{Salcedo-Sanz2005, Wang2015}.


The increased flexibility of modern satellite payloads, including the ability to dynamically reallocate frequency resources, has been widely studied for GEO systems. Most works focus on allocating frequency, bandwidth, or both, for which greedy algorithms \cite{Kiatmanaroj2012a}, neural networks \cite{Ortiz-Gomez2021} and deep reinforcement learning \cite{Hu2020} have been proposed. Other works also optimize the power allocation, presenting heuristic methods \cite{Alberti2010}, metaheuristics \cite{Cocco2018, Paris2019}, and convex approximation \cite{Abdu2021a}, usually in combination with iterative procedures to deal with the complexity of the problem. Limited works specifically consider mobile users \cite{Abe2018, Hu2020, Abdu2022}, modeling them as demand variations within the beams. The proposed solutions either generate frequency assignments given known demand distributions or dynamically reallocate resources in real-time based on demand changes. In that sense, they do not consider information that can be known before operations, such as mobile users' trajectories, to generate a frequency allocation valid for a given time interval.


In contrast to GEO systems, frequency assignment for NGSO constellations has been severely understudied, and, to the best of the authors knowledge, no previous works directly consider use cases with mobile users. In \cite{Kisseleff2019}, the author proposes an optimization tool for LEO systems that can achieve precise traffic matching by jointly optimizing multiple system parameters using evolutionary algorithms and neural networks. In \cite{PachlerdelaOsa2021}, the authors present a user-centric beam coverage and a heuristic frequency assignment algorithm for LEO constellations. In \cite{Garau-Luis2022} the authors implement an iterative dynamic frequency assignment algorithm that can prioritize different operational requirements and evaluate their solution in an MEO constellation. Using a similar approach, the allocation of multiple satellite resources has been considered in \cite{pachler22b}. However, these methods have been evaluated in static scenarios without user movement. Thus, it is unclear if they can be adapted for the demand characteristics of mobile users.


While broadband services for mobile users has not been studied, mobile users have been included in studies focusing on phone communications in LEO constellations \cite{DelRe1993, Maral1998, Zheng2020}. Most works neglect terminal mobility by considering satellite-fixed cell coverage, with beam footprints moving relative to the ground. Inspired by terrestrial networks, they focus on fixed and dynamic channel assignment and handover management to reduce call blocking and dropping rates.


Given the growing demand for broadband connectivity on the move, spectrum management strategies need to leverage the flexibility of modern NGSO satellites to provide service robustness against dynamic user bases. Mobile users add a layer of complexity to the frequency assignment problem due to their contribution to the geographical and temporal unbalance of traffic demand, which constitutes a new source of uncertainty. For this reason, proactively including user information forecasts when optimizing the frequency plan and reactively reallocating resources in real-time can be a key efficiency driver. The presented studies fail to study the complexity added by mobile users to the frequency assignment problem, and none considers including demand forecasts, user position forecasts, or uncertainty considerations to improve satellite resource utilization.

\subsection{Paper objectives}



To close this research gap, the objective of this paper is to present a dynamic frequency assignment framework based on linear programming that can operate in the presence of mobile users. The framework relies on a two-stage process to account for the complexity and uncertainty introduced by these users. The first stage consists of proactive long-term planning, and the second reallocates resources in real-time. This synergy between long-term and short-term frequency assignments has not yet been exploited for communication satellite systems. Our method leverages the flexibility of current satellite systems by including full frequency reuse using spatial separation and different polarizations, dynamic bandwidth assignment, and operational aspects such as the inclusion of gateway restrictions, all of which help bridge the research gap between technical research and operational constraints.



\subsection{Paper structure}


The remainder of the paper is organized as follows: Section \ref{sec:formulation} formulates the frequency assignment problem for NGSO constellations with the implications introduced by mobile users, Section \ref{sec:methods} describes the proposed frequency assignment method with its proactive and reactive stages to operate under uncertainty, Section \ref{sec:results} shows the results of applying the method to different use cases, and finally Section \ref{sec:conclusions} remarks the conclusions of this work and possible future research directions.

\section{The frequency assignment problem}
\label{sec:formulation}

\begin{figure*}[tb]
    \centering

    \begin{subfigure}[b]{0.61\textwidth}
        \centering
        \includegraphics[width=0.98\textwidth]{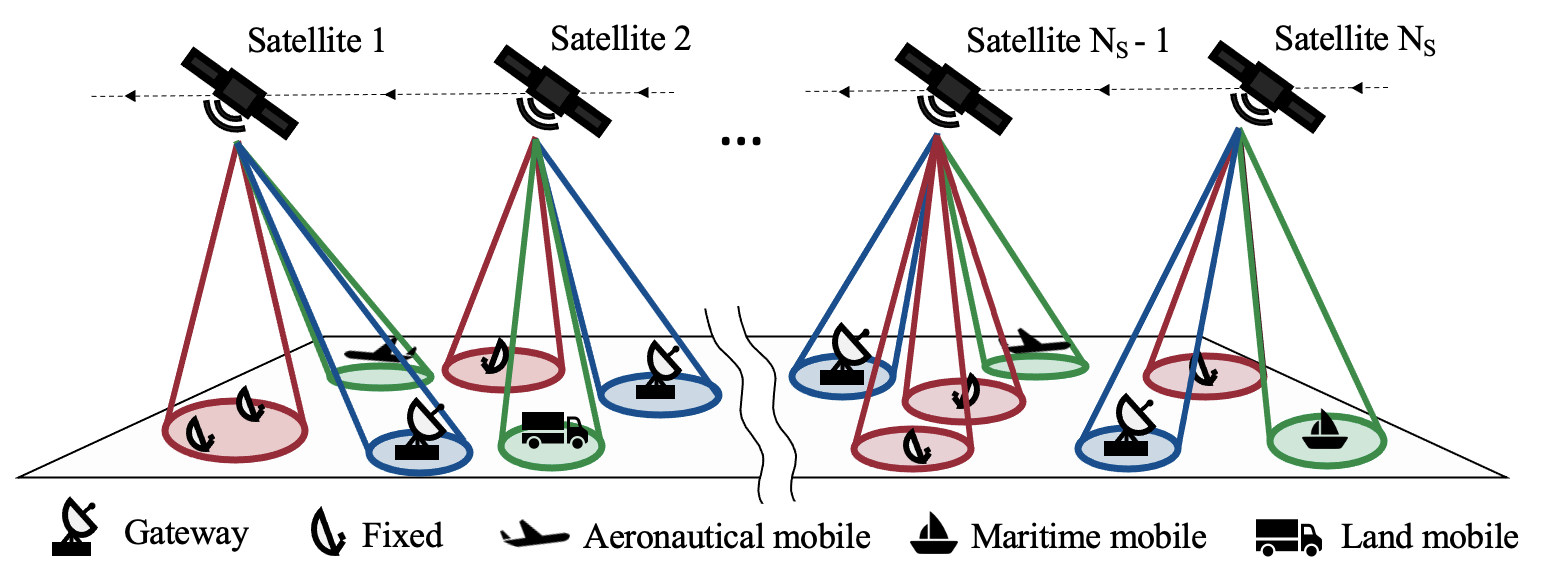}
        \hfill
    \end{subfigure}
    \hfill
    \begin{subfigure}[b]{0.38\textwidth}
        \hfill
        \includegraphics[width=0.98\textwidth]{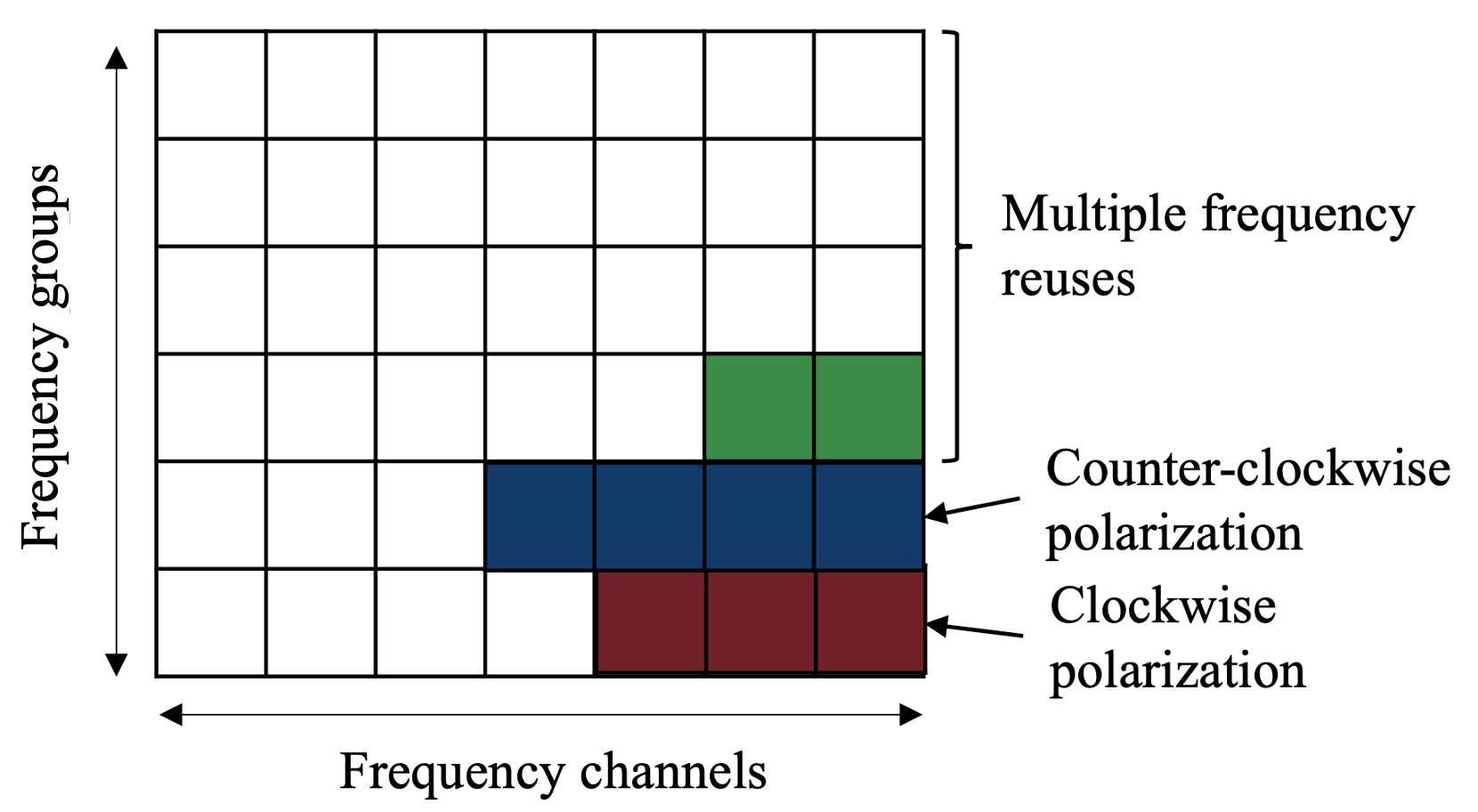}
    \end{subfigure}
    \caption{NGSO constellation plane with $N_S$ satellites connecting fixed and mobile users to gateways and frequency assignment representation in the form of a grid with $N_{r}\cdot N_p$ rows (frequency groups) and $N_{ch}$ columns (frequency channels). In this example, $N_{r} = 3$, $N_p = 2$, and $N_{ch} = 7$. In this case, the allocation in blue is defined by $f=4$, $b=4$, $g=1$, $p=2$.}
    \label{fig:formulation}
\end{figure*}

In this work, we pose the frequency assignment as an optimization problem and extend the formulation proposed in \cite{Garau-Luis2022}. A key difference is that, while the original formulation disregards the temporal dimension, we frame the problem as generating a frequency assignment that considers changes in the user distribution occurring during a time period $t\in[0,T]$.

\subsection{System description}

We consider an NGSO constellation with $N_S$ identical multibeam satellites that connect fixed and mobile users to fixed gateways, which are ground stations that transmit data between satellites and terrestrial networks (see Figure \ref{fig:formulation}). The set of users and gateways are connected using $N_B$ downlink beams. While fixed users can be defined using a single static position $p_u$ and require service at all times, mobile users require service at different positions $p_u(t)$ during specific time periods $t\in[t_{start,u},t_{end,u}]$, where $t_{start,u}$ and $t_{end,u}$ are the start and end service times, respectively. We assume that multiple fixed users can be served by one fixed beam, whereas each mobile user is assigned to a beam that follows them throughout their trajectory. Each beam is powered by a single satellite at any point in time. Satellite handover operations, defined as the change of the satellite servicing a beam, occur due to the orbital dynamics of NGSO constellations and the movement of mobile users.

We assume all satellites have access to the same frequency spectrum, divided into $N_{ch}$ equal-bandwidth channels. The payload can reuse frequency up to $N_r$ times and handle $N_p$ polarizations, allowing close beams to use the same frequency without incurring additional interference.
A combination of specific reuse and polarization is referred to as a frequency group.

Generating a frequency plan involves assigning, to every beam $i$, a set of adjacent frequency channels for each time instant this beam is active (i.e., continuously for fixed users, or between $t_{start,i}$ and $t_{end,i}$ for mobile). This allocation is defined by an initial frequency channel $f_i(t)$, the number of consecutive channels $b_i(t)$, a frequency reuse $r_i(t)$, and a polarization $p_i(t)$. Although it might not be preferred, we assume that the assignment of beams can be changed at any time. Figure \ref{fig:formulation} shows the representation of the frequency assignment in the form of a grid with $N_r \cdot N_p$ rows (frequency groups) and $N_{ch}$ columns (frequency channels).

As detailed in Appendix \hyperref[sec:power]{A}, we can compute the minimum $b_{min,i}$ and the maximum $b_{max,i}$ number of frequency channels for each beam $i$ so that the communication link is feasible. Taking that into account, the boundaries and domains of the frequency assignment variables are:

\begin{equation}
    \label{eq:variables}
    \begin{split}
        &1 \leq f_i(t) \leq N_{ch} \\
        &f_i(t)+b_i(t)-1 \leq N_{ch} \\
        &b_{min,i} \leq b_i(t) \leq b_{max,i} \\
        &1 \leq g_i(t) \leq N_r \\
        &1 \leq p_i(t) \leq N_p \\
        &f_i(t), b_i(t), g_i(t), p_i(t) \in \mathbb{Z}^{+} \\
        & \forall\ i \in \{1,...,N_B\}\\
        & \forall\ t \in [0,T]
    \end{split}
\end{equation}



\subsection{Handover and interference constraints}

The frequency assignment in NGSO constellations is subject to two main types of constraints: handover and interference restrictions. We first explain the former.

\begin{figure}[ht]
    \centering
    \includegraphics[width=0.49\textwidth]{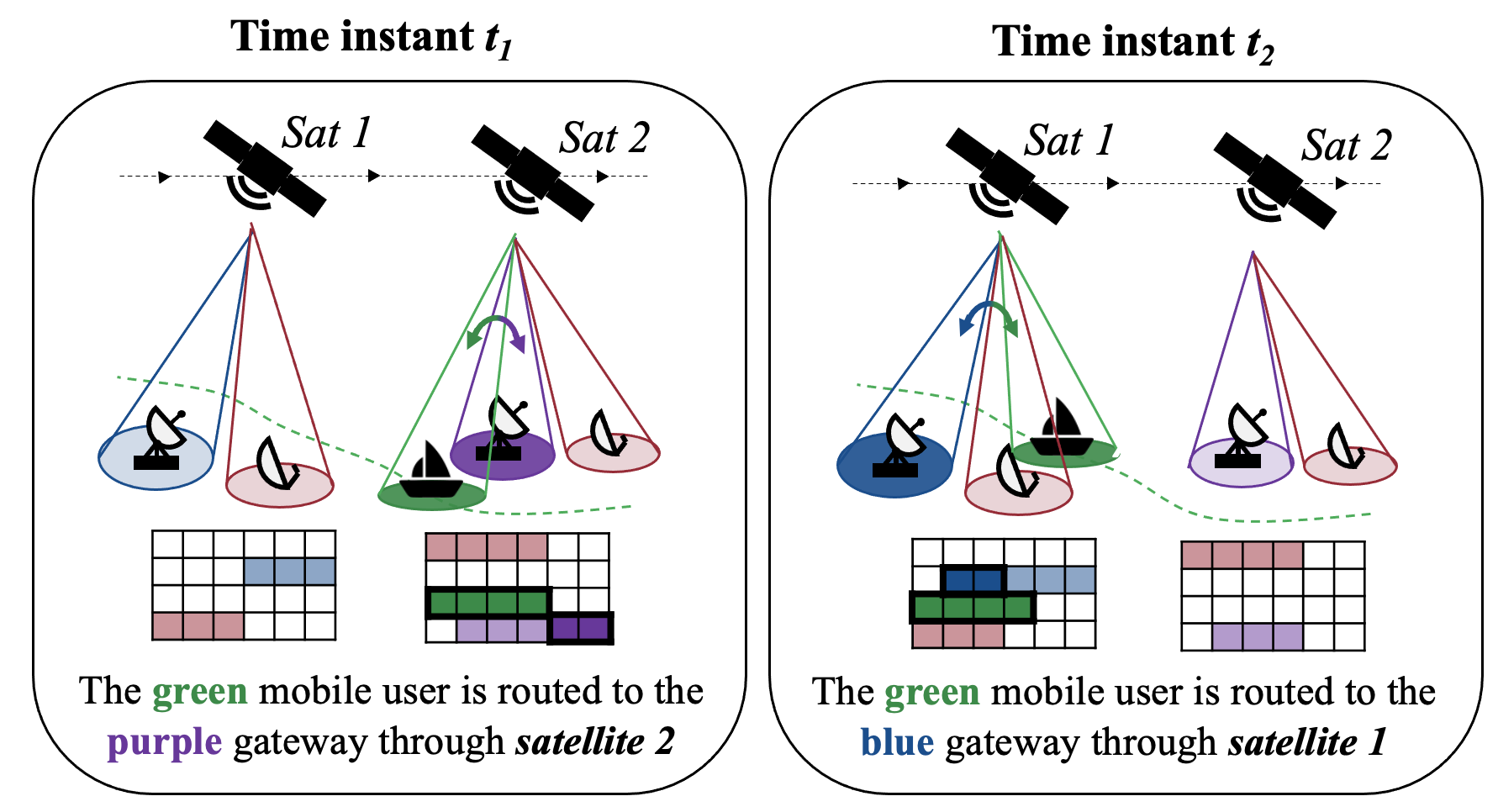}
    \caption{User distribution and frequency assignment of two satellites at two different instances. Contrary to fixed users, the change in position of the mobile user in green prompts a change in the frequency allocation (highlighted in a darker color).}
    \label{fig:f_temp_changes_1}
\end{figure}

While a beam is being served by a satellite, we must ensure that no other beam uses the same resources. To account for this, we define a handover constraint for all beams served by the same satellite, which prevents them from being assigned overlapping frequency channels and the same frequency group. These constraints are encoded for each pair of beams $(i,j)$ using the binary variables $\beta_{ij}(t)$. If two beams are assigned to the same satellite at time $t$, then $\beta_{ij}(t)=1$ ($\beta_{ij}(t)=0$ otherwise). These constraints allow capturing the handovers from both the dynamics of NGSO constellations and those derived from the change in position of mobile users. As an example, in Figure \ref{fig:f_temp_changes_1}, the terminals covered by the green and blue beams are not simultaneously in the field of view of the same satellite around time $t_1$ ($\beta_{green, blue}(t_1)=0$), meaning that they could use the same frequency resources. However, due to the change in position of the mobile user in the green beam, they are served by the same satellite around time $t_2$ ($\beta_{green,blue}(t_2)=1$), which forbids them from using the same resources.

In addition to satellite limitations, beams whose footprints are geographically close are susceptible to cause interference with one another if they use the same polarization and overlapping frequency channels. To account for this, we define an interference constraint for all geographically close beams. These constraints are encoded for each pair of beams $(i,j)$ using the binary variables $\alpha_{ij}(t)$ for each time $t$, which depend on the position of the two beams, $p_i(t)$ and $p_j(t)$. As proposed in \cite{PachlerdelaOsa2021}, we consider that two beams are susceptible to interference, with $\alpha(t)=1$ ($\alpha(t)=0$ otherwise), if the angular separation (measured from the satellites) between their footprint centers $\delta(p_i(t),p_j(t))$ is lower than a certain threshold $\delta_{min}$. In contrast to fixed user bases, the change in position of mobile users is the major cause of the temporal change of interference constraints. As illustrated in Figure \ref{fig:f_temp_changes_1}, the footprint of the green beam is near that of the purple beam at time $t_1$ ($\alpha_{green,purple}(t_1)=1$, $\alpha_{green,red}(t_1)=0$), whereas at time $t_2$ it is close to that of the red beam  ($\alpha_{green,purple}(t_0)=0$, $\alpha_{green,red}(t_2)=1$).

To avoid invalid assignment of resources while accounting for interference between beams, we need to impose the following handover constraints:
\begin{equation}
    \begin{split}
        \label{eq:intra}
        &\beta_{ij}(t) = 0 \text{ or } g_i(t)\neq g_j(t) \text{ or } p_i(t)\neq p_j(t)\\
        &\text{or } f_i(t) + b_i(t) \leq f_j(t) \text{ or } f_j(t) + b_j(t) \leq f_i(t) \\
        &\forall\ i,j \in \{1,...,N_B\},\;i\neq j\\
        &\forall\ t \in [0,T]
    \end{split}
\end{equation}
And the following interference constraints:
\begin{equation}
    \begin{split}
        \label{eq:inter}
        &\beta_{ij}(t) = 0 \text{ or } \alpha_{ij}(t) = 0 \text{ or } p_i(t)\neq p_j(t)\\
        &\text{or } f_i(t) + b_i(t) \leq f_j(t) \text{ or } f_j(t) + b_j(t) \leq f_i(t) \\
        &\forall\ i,j \in \{1,...,N_B\},\;i\neq j\\
        &\forall\ t \in [0,T]
    \end{split}
\end{equation}

For a detailed formulation of the constraints, including their linearization, the reader is referred to \cite{Garau-Luis2022}.

\subsection{Objective function}

Since power is usually one of the main limiting factors onboard a satellite, frequency plans with lower power consumption are inherently more attractive for the operators. Thus, we define power as the primary driver of optimality, transforming the problem into:

\begin{equation}
    \label{eq:problem}
    \begin{aligned}
        \min_{}\quad & \sum_{b=1}^{N_B} \int_{t_{start,i}}^{t_{end,i}}P_i(f_i(t), b_i(t))  \,dt \\
        \textrm{s.t.} \quad
                     & \textrm{Eq. (\ref{eq:variables}})                                        \\
                     & \textrm{Eq. (\ref{eq:inter}})                                            \\
                     & \textrm{Eq. (\ref{eq:intra}})                                            \\
    \end{aligned}
\end{equation}


where $P_i(f_i(t), b_i(t))$ is the radio-frequency transmission power needed to serve beam $i$ at time $t$.
Detailed steps regarding its calculation can be found in Appendix \hyperref[sec:power]{A}.


\subsection{Mobile users and uncertainty}
\label{sec:mobile_users_uncertainty}

Besides the time-dependency of the constraints, generating a frequency plan for mobile users has the additional challenge of planning with uncertainty since information about their position comes at different times and with different accuracies. Therefore, efficiently solving the problem requires both taking advantage of the already known information and adapting to new information. For example, satellite operators might know the expected flight departure time, origin, and destination several days in advance, whereas the exact trajectory that aircraft will follow will not be known until moments before departure or even when they are en route. Other users, such as trucks, might request service in specific locations without prior notice; thus, their service times and positions are hard to predict.


When new information about a user is known, additional restrictions can arise in the frequency assignment. For example, unexpected flight delays alter the time interval when resources are required from satellites and gateways. Trajectory changes entail that users will not be at the expected positions, which, together with delays, can add interference and handover constraints with other users. Due to these situations, the beams might have a lower angular separation than anticipated or need to be assigned to the same satellite when it was not scheduled. Since uncertainty can invalidate the frequency allocation, it must be accounted for when providing service to mobile users by proactively protecting the frequency assignment with conservative constraints or reactively modifying it. Our method is framed around this whole operational context and leverages both prior information on mobile users to improve frequency assignment before operations and an efficient technical approach to address real-time changes due to unexpected events. We describe it in the following section.

\section{Method overview}
\label{sec:methods}

\begin{figure*}[t]
    \centering
    \includegraphics[width=0.8\textwidth]{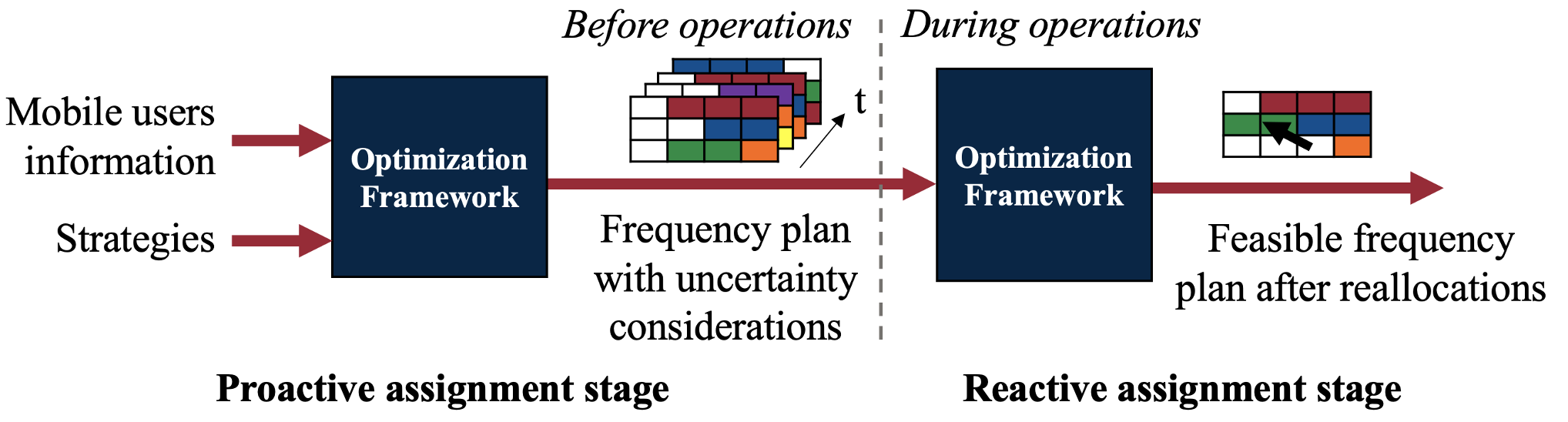}
    \caption{Proposed frequency assignment framework, consisting of a proactive assignment stage that generates a baseline plan accounting for uncertainty before operations, and a reactive assignment stage that re-optimizes the plan during operations when new information is available.}
    \label{fig:m_framework}
\end{figure*}


Frequency assignment frameworks should not only seek to find optimal or quasi-optimal solutions but also be robust in dynamic and uncertain environments, especially when real-time operation is costly. To that end, we propose a two-phase frequency allocation method divided into a proactive and a reactive assignment stage. As illustrated in Figure \ref{fig:m_framework}, before the beginning of operations, we leverage available prior information and generate a baseline plan accounting for uncertainty; this constitutes the proactive stage. Later, during operations, this baseline plan is re-optimized when new information is available (e.g., an unexpected event occurs or users with no prior information enter the system); this corresponds to the reactive stage. Our framework overcomes mobile user uncertainty by leveraging conservative decision-making in the proactive stage and a real-time re-optimization procedure in the reactive stage.


This two-step approach has been successful in similar applications. In particular, it has shown promising results in the resource-constrained project scheduling (RCPS) problem, which consists of assigning temporal resources to a set of activities over time \cite{Davari2019}.
The sources of uncertainty in the RCPS, such as uncertain activity duration, unexpected addition of new activities, or the need to change the start times, cause disruptions similar to those in the frequency assignment problem \cite{Habibi2018}. We now explain the details of our method.

\subsection{Proactive frequency assignment stage}


The goal of the proactive planning stage is to construct, before the beginning of operations, a baseline frequency plan that considers both fixed and mobile users and can be operated during a specific time period $t\in[0,T]$. We denote by $U_{tot}$ the set of all users that will require service during this time interval. Out of those, we assume there is information on a subset $U_{info}$, with $|U_{info}| \leq |U_{tot}|$, available beforehand; this includes data rate demands, location, and schedules and trajectories in the case of mobile users. For the users in $U_{info}$, this information is leveraged to compute beam placements, beam routings, and handover and interference constraints, which are inputs to our frequency assignment algorithm.

Our method assumes each mobile user is assigned to a single beam, whereas fixed users might be grouped and assigned to the same beam. The complete footprint layout is then used to compute the routing and handover schedules, which are necessary in the case of NSGO constellations. We leave the specific methods used to carry out these operations outside this paper's scope; we assume the footprint layouts and the NSGO schedule are provided beforehand. Then, knowing the handover schedule and the footprint layout allows for computing the constraints between beams that will take place during operation. With this information, a full frequency plan is defined for all users in $U_{info}$ for the time frame considered. In the following lines, we first outline the technical approach in the absence of uncertainty and then introduce robustness mechanisms that account for it.

\subsubsection{Mobile users without uncertainty}

Our method stems from the work in \cite{Garau-Luis2022}, which presents a frequency assignment algorithm based on Linear Programming (LP)---we refer to the original paper for the complete details on its formulation. The original method is only defined for fixed users, which allows computing a single plan regardless of the time horizon since handover and interference constraints remain static. In our case, constraints are time-dependent due to the change in position of mobile users.

Let $N_B$ be the total number of beams in the problem (which might be smaller than $|U_{info}|$ given we group fixed users).
To make use of the LP method from \cite{Garau-Luis2022}, we define the inter-group and intra-group restrictions sets $\mathcal{R}_E$ and $\mathcal{R}_A$---introduced in the original paper---as:
\begin{equation}
    \begin{aligned}
        \label{eq:intrainter}
         & (i, j) \in \mathcal{R}_A,\quad \text{if } \underset{\begin{subarray}{c}
                                                                       t\in[0,T]
                                                                   \end{subarray}}{\max}\,\alpha_{ij}(t) = 1 \\
         & (i, j) \in \mathcal{R}_E, \quad \text{if } \underset{\begin{subarray}{c}
                                                                        t\in[0,T]
                                                                    \end{subarray}}{\max}\,\beta_{ij}(t) = 1
    \end{aligned}
\end{equation}
where $\mathcal{R}_E$ and $\mathcal{R}_A$ correspond to the sets of beams that hold, at any moment in time, an interference or handover restriction, respectively. The variables $\alpha$ and $\beta$ are those defined in the previous section, and $T$ is the length of the considered time interval. Note that our approach considers the worst-case scenario in the trajectory of mobile users, eliminating the temporal dependency of the frequency and handover constraints. This way, we are able to extend the original LP method designed only for fixed users to define complete frequency plans for both fixed and mobile users.

\subsubsection{Mobile users with uncertainty}

As mentioned in the introduction of this paper, uncertainty is intrinsic to long-term decision-making in the presence of mobile users; their schedules and trajectories might differ from those planned for multiple reasons.
While we propose adding a reactive stage during operations to address unexpected changes, we find that already accounting for uncertainty during the proactive stage can reduce the burden of reallocating during operations.
In those cases, minor delays or deviations can significantly impact the frequency plan by changing the constraints in real-time.

Proactive strategies have been widely used in the RCPS problem \cite{Davari2019}, such as extending activity duration based on their statistics or spacing out tasks based on a factor adjusted by simulation. In our case, this protection can consist of assuming larger user service times and areas to minimize disruptions. Accordingly, we propose three different reactive strategies. Before presenting them, we redefine the problem variables $\alpha$, $\beta$ as follows:

\begin{itemize}[itemsep=0em]
    \item First, we define $\mathcal{P}_i(t)$ as the set of possible positions of beam $i$ at time $t$. This area can be computed as function of the expected position $p_u(t)$ of a mobile user $u$, and a parameter $\gamma$ that defines the magnitude of the possible deviations, and thus its size: $\mathcal{P}_i(t)=f(p_u(t),\gamma)$.
    \item Next, we define the minimum angular separation between the sets of possible positions of beam $i$ at time $t_i$ and beam $j$ at time $t_j$ as:
          \begin{equation}
              \Delta(\mathcal{P}_i(t_j), \mathcal{P}_j(t_j)) =  \underset{\begin{subarray}{c}
                      p_i\in\mathcal{P}_i(t_i)\\
                      p_j\in\mathcal{P}_j(t_j)
                  \end{subarray}}{\min} \delta(p_i,p_j)
          \end{equation}
          where $\delta(p_i,p_j)$ is the angular separation between two beams at positions $p_i$ and $p_j$, respectively. Note that we now consider the worst case between two sets of positions at different time instants, whereas our previous definition considered two positions at the same time instance $\delta(p_i(t),p_j(t))$.
    \item Consequently, we define $\alpha_{ij}(t_i, t_j)$ to be 1 if $\Delta(\mathcal{P}_i(t_j), \mathcal{P}_j(t_j)) \leq \delta_{min}$, checking whether beam $i$ at time $t_i$ could interfere with beam $j$ at time $t_j$.
    \item Similarly, $\beta_{ij}(t_i, t_j)$, encodes the handover restrictions when beams have a set of possible positions for a given time $t$.
\end{itemize}

Using this notation, the three proactive strategies our method uses are:

\textbf{S1. Larger service times}\quad When computing interference and handover constraints, we assign resources during longer service times by considering that users might be delayed up to $t_d$. To that end, we redefine equations (\ref{eq:intrainter}) for $\mathcal{R}_E$ and $\mathcal{R}_A$, such that $(i, j) \in \mathcal{R}_E$ if:
\begin{equation}
    \underset{\begin{subarray}{c}
            t\in[0,T]
        \end{subarray}}{\max}\,\pqty{ \underset{\tau_1,\tau_2\in[0,t_d]}{\max}\, \alpha_{ij}(t + \tau_1, t + \tau_2)} = 1
\end{equation}
and $(i, j) \in \mathcal{R}_A$ if:
\begin{equation}
    \underset{\begin{subarray}{c}
            t\in[0,T]
        \end{subarray}}{\max}\,\pqty{ \underset{\tau_1,\tau_2\in[0,t_d]}{\max}\, \beta_{ij}(t + \tau_1, t + \tau_2)} = 1
\end{equation}

\textbf{S2. Larger interference threshold}\quad To make interference constraints more conservative, we update the distance threshold for $x_{min}\cdot\delta_{min}$, with $x_{min} > 1$.

\textbf{S3. Larger operational areas}\quad We can define bounded regions where it is likely to find the users at a given time $\mathcal{P}_i(t_i,\gamma)$, where $\gamma$ allows adjusting its size. This allows specifying tailored operational areas that directly impose interference constraints on beams that traverse them. For example, if we know that a cruise ship will operate in the Caribbean region $\mathcal{P}$, we can set $\mathcal{P}_i(t_i)=\mathcal{P}$ for the beam serving it.

Using these strategies when computing the baseline frequency plan for users in $U_{info}$ allows the system to capture possible delays and trajectory changes in real-time without recomputing the frequency assignment. However,
since not all uncertainty can be captured,  we leverage a reactive stage to make changes in real-time; we introduce it in the following subsection.


\subsection{Reactive assignment stage}



In contrast to the proactive stage, the reactive assignment stage revises and re-optimizes the baseline plan during its execution period $t\in[0,T]$, when new information about the users is known. The capability of modifying the frequency plan is crucial for two reasons:
\begin{enumerate}[itemsep=0em]
    \item The baseline plan is generated based on the information on a subset of users $U_{info}$, with $|U_{info}|\leq|U_{tot}|$, since some users might request service in a specific location without prior notice.
    \item The information known about mobile users in $U_{info}$ changes over time. For example, operators might have new information about the exact route and departure time, which was unknown when generating the baseline plan.
\end{enumerate}




Reactive strategies have been studied in the RCSP problem \cite{Habibi2018}, 
with rescheduling polices that can cope with the insertion of unanticipated activities into a given baseline schedule, which in our case would be similar to adding users in $U_{tot}$ that are not in $U_{info}$. 
These strategies also include having multiple identical sets of resources kept in standby or scheduling back-up tasks, which in our case can consist of reserving part of the spectrum or having back-up frequency assignments.

Motivated by these approaches, we make use of three different strategies that consist of reserving resources when generating the baseline plan to use them in case of unexpected events:



\textbf{S4. Reserving additional channels}\quad We can reserve adjacent frequency channels for beams serving mobile users in $U_{info}$. Since the channels are adjacent to the beam's allocation, the problem's dimensionality does not increase. We reserve $x_{ch}\cdot b_{min,i}$ channels for each beam $i$, thus $(x_{ch}+1)\cdot b_{min,i}\leq b_i \leq b_{max,i} + x_{ch}\cdot b_{min,i}$, with $x_{ch}\geq1$. Note that $b_{min,i}$ is the minimum required number of channels.

\textbf{S5. Reserving additional slots}\quad Having a set of backup assignments (slots) for beams serving mobile users in $U_{info}$. If new constraints invalidate the original assignment of a beam, it can be reallocated to one of those slots, which have been optimized before operations.
We propose assigning $x_{slots}$ slots of $b_{min,i}$ channels to each beam $i$, with $x_{slots}\geq1$.
In contrast to the previous strategy, reserving non-adjacent channels increases the flexibility of the allocation and the chance of reallocating successfully, at the expense of increasing the number of assignments to be performed.

\textbf{S6. Reserving spectrum}\quad Reserving a fraction of frequency channels that cannot be used when computing the baseline frequency plan but serve as emergency channels when needing to reallocate users in real-time. This can be implemented by redefining variables in (\ref{eq:variables}) and setting $1+\lceil x_{spec} \cdot N_{ch}\rceil \leq f_i$, for all $i \in\{1,...,N_B\}$, with $0< x_{spec} < 1$.

The use of these strategies when generating the baseline frequency plan for users in $U_{info}$ allows leveraging reserved resources when re-computing the assignment in real-time when all $U_{tot}$ users come into play.

\section{Results}
\label{sec:results}

This section presents the metrics used to evaluate our method and introduces the constellation and user models. Next, we validate our approach by assuming a scenario without uncertainty. Finally, we analyze the presented framework under uncertainty using proactive and reactive strategies.

\subsection{System metrics}

To compare the different strategies, we have to define the evaluation criteria. Ideally, we would like to identify which strategy or set of strategies are the ones that provide the greatest value for our system, defined as a benefit at cost. For this purpose, we detail a cost metric and a performance metric.

\textit{Cost metric}---For the cost metric, we use the total power consumption, as power is usually one of the limiting factors onboard a spacecraft. Thus, frequency plans with lower power consumption are inherently more attractive \cite{Garau-Luis2022}. To eliminate the time dependency, the metric is defined as the average total power consumption $P$ of the constellation required to serve the demand of each beam:
\begin{equation}
    P = \frac{1}{T}\sum_{i=1}^{N_B}\int_{t_{start,i}}^{t_{end,i}}  P_i(f_i(t), b_i(t))\,dt
\end{equation}
where $T$ is the length of the considered period, $P_i$ is the power consumption of beam $i$, which is calculated as detailed in Appendix \hyperref[sec:power]{A}.

\textit{Performance metric}---Since our method allows deactivating beams that cannot be assigned any resources, the power consumption alone does not reflect how effectively we overcome uncertainty. For this reason, we make use of the number of successfully served users $N_{served}$ as the performance metric. Conservatively, if any beam connecting a user to a gateway (or gateways) violates a constraint and their reallocation is unsuccessful, these beams are deactivated, and the user is considered not to be served.



As an additional metric, since operators might prefer to minimize the number of reconfigurations in real-time due to unanticipated events, we define $N_{realloc}$ as the number of reallocations that happen during operations.


\subsection{Experimental setup}
\label{sec:setup}
We simulate scenarios with fixed, aeronautical, maritime, and land mobile users, using data from open source datasets (see Appendix \hyperref[sec:users]{B}). We use a downsized MEO constellation similar to O3b mPower, \cite{FederalCommunicationsCommission2020}
with the parameters specified in Table \ref{tab:parameters}. Fixed users are first grouped into a set of beams using the algorithm proposed in \cite{PachlerdelaOsa2021}. Users are assumed to be routed to the closest gateway and the closest satellite.

\begin{table}[b!]
    \centering
    \footnotesize
    \begin{tabular}{lccc}
        \hline
        \rowcolor{lightbg}
        Parameter                             & Value \\
        \hline
        Number of satellites $N_s$            & 7     \\
        Number of frequency channels $N_{ch}$ & 80    \\
        Number of frequency reuses $N_r$      & 8     \\
        Number of polarizations $N_p$         & 2     \\
        Frequency band                        & Ka    \\
        \hline
    \end{tabular}
    \caption{Constellation parameters.}
    \label{tab:parameters}
\end{table}

We consider two scenarios with 245 and 330 users ($|U_{tot}|$), as well as two levels of uncertainty (low and high). The level of uncertainty is determined based on the magnitude of the uncertainty in the position and delay of aeronautical mobile users in $U_{info}$ and the fraction of land mobile users not included in $U_{info}$ (see Appendix \hyperref[sec:users]{B}). We assume that the position of maritime users is accurately known a priori, and while they do not introduce anticipated events, they contribute to the changes in demand distribution. The users are concentrated in a specific region of the globe to reach a high number of handover and interference restrictions without the need to optimize for a global user basis.

\subsection{Frequency assignment without uncertainty}

In this first experiment, we evaluate our method in scenarios without uncertainty, i.e., $U_{info} = U_{tot}$, to assess that it is able to produce frequency assignments for both fixed and mobile users. Since all information about users is known a priori, the baseline plan remains feasible and does not require using the reactive assignment stage or strategies S1-S6.

\begin{table}[t!]
    \footnotesize
    \centering
    \begin{tabular}{cccc}
        \hline
        \rowcolor{lightbg}
        $N_{users}$ & $N_{beams}$ & $N_{served}/N_{users}$ & $P/P_{sat}$   \\
        \rowcolor{lightbg}
                    & Avg. (SD)   & Avg. (SD)              & Avg. (SD)     \\
        \hline
        245         & 730 (19)    & 1.000 (0.000)          & 0.203 (0.050) \\
        330         & 938 (19)    & 1.000 (0.000)          & 0.515 (0.122) \\
        \hline
    \end{tabular}
    \caption{Results of the frequency assignment (no uncertainty, $U_{info}=U_{tot}$). $N_{beams}$ is the number of beams (mobile users might connect to multiple gateways), $N_{served}/N_{users}$ is the fraction of successfully served users, and $P/P_{sat}$ is the power consumption normalized to the estimated power capabilities of a satellite. We indicate the average and standard deviation (SD) from 30 runs.}
    \label{tab:exp1_results}
\end{table}
\begin{figure}[!ht]
    \centering
    \includegraphics[width=1.0\linewidth]{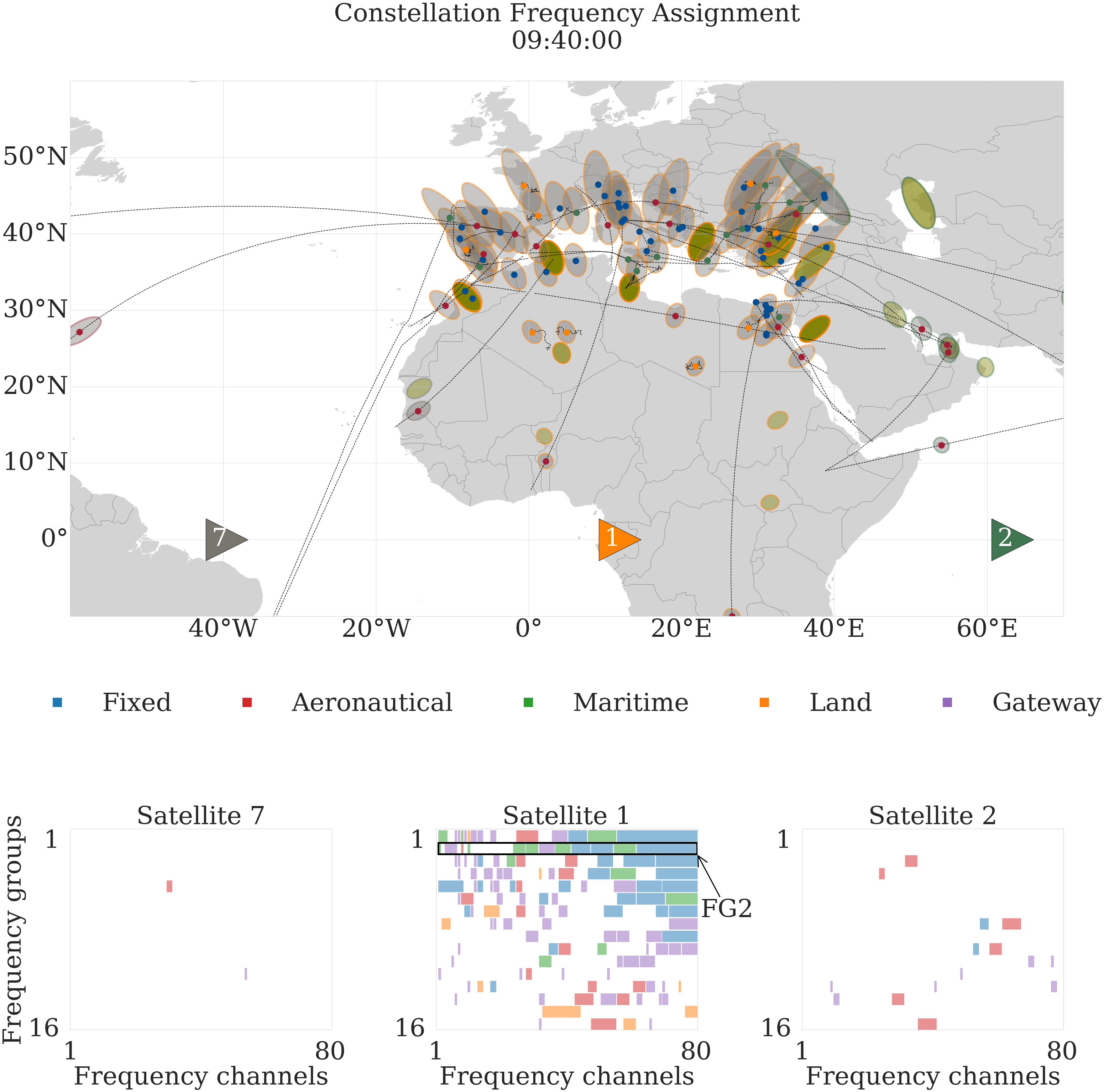}
    \caption{Frequency assignment (no uncertainty, $U_{info}=U_{tot}$) at time 9h40min. The upper part shows the users' positions along their trajectories, where the shaded areas are the beam footprints. The lower part shows the frequency assignment, highlighting the frequency group plotted in Figure \ref{fig:r_rg_plan}.}
    \label{fig:r_constellation_plan}
\end{figure}
\begin{figure}[!ht]
    \centering
    \begin{subfigure}[t]{1.0\linewidth}
        \centering
        \includegraphics[width=1\linewidth]{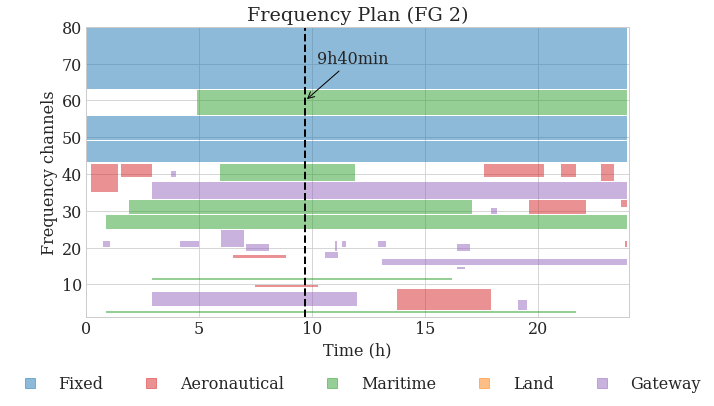}
    \end{subfigure}
    \newline
    \begin{subfigure}[t]{1.0\linewidth}
        \centering
        \includegraphics[width=0.9\linewidth]{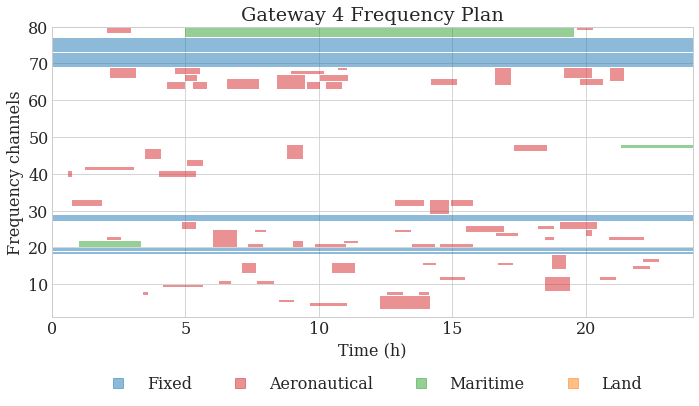}
    \end{subfigure}
    \hfill
    \caption{Baseline frequency plan with no uncertainty ($U_{info}=U_{tot}$) from the perspective of the satellites (frequency group 2) and one of the gateways. The horizontal axis shows the simulation time $t\in[0,T]$, with $T=24$h, and the vertical axis shows the frequency channels, with $N_{ch}=80$. The dashed vertical line indicates the instance 9h40min, for which the constellation assignment is plotted in Figure \ref{fig:r_constellation_plan}.}
    \label{fig:r_rg_plan}
\end{figure}

Table \ref{tab:exp1_results} shows results of 30 runs simulating different user distributions and thus different user positions and service times. As observed, our framework can meet all the demand in scenarios with 245 and 330 users, proving that it can generate a frequency plan before operations based on accurate information. Figure \ref{fig:r_constellation_plan} shows the user distribution and the frequency assignment at a specific instance of one of the runs. Since users are concentrated in a small region, most of them are served by satellite 1 . The unassigned frequency resources in that satellite might be occupied by beams that will
undergo or have recently undergone a handover or by beams that are not active at this instant.

Figure \ref{fig:r_rg_plan} shows one frequency group of the generated baseline frequency plan from the perspective of the satellites and from that of a gateway. The baseline plan assigns the same resource to multiple beams at different times while respecting interference and handover constraints (there are no overlapping assignments).

As mentioned, since we generate a baseline plan with complete information about the users ($U_{info}=U_{tot}$), the reactive stage does not carry out real-time reallocations ($N_{realloc}=0$). In the next experiment, we discuss how our framework can operate under uncertainty.

\subsection{Frequency assignment under uncertainty}

\begin{table*}[t]
    \centering
    \footnotesize
    \begin{tabular}{crcccrcccc}
        \hline
        \rowcolor{lightbg}
        Experiment config.
        & \multicolumn{4}{c}{Proactive strategy (S1-S3)}
        & \multicolumn{5}{c}{Reactive strategy (S4-S6)} \\
        \hline
        $A$ & \multicolumn{4}{c}{--} & \multicolumn{5}{c}{--} \\
        \hline
        $B$ & \multicolumn{4}{c}{--} & \multicolumn{5}{c}{$x_{ch}=1$} \\
        \hline
        $C$ & \multicolumn{4}{c}{--} & \multicolumn{5}{c}{$x_{slots}=1$} \\
        \hline
        $D_1$, $D_2$, $D_3$, $D_4$ 
        & \multicolumn{4}{c}{--} 
        & $x_{spec}=$ & 0.05 & 0.10 & 0.15 & 0.20
        \\
        \hline
        \multirow{2}{*}{$E_1$, $E_2$, $E_3$}
        & $t_d=$     & $p_{50th}$ & $p_{75th}$ & $p_{95th}$
        & \multicolumn{5}{c}{\multirow{2}{*}{--}}
        \\
        & $x_{min}=$ & 1.15       & 1.30       & 1.45
        & \\
        \hline
        \multirow{2}{*}{$F_1$, $F_2$, $F_3$}
        & $t_d=$      & $p_{50th}$ & $p_{75th}$ & $p_{95th}$
        & \multicolumn{5}{c}{\multirow{2}{*}{--}}
        \\
        & $\gamma=$   & $p_{50th}$ & $p_{75th}$ & $p_{95th}$ \\
        \hline
        $G_1$, $G_2$, $G_3$
        & $t_d=$      & T          & T          & T
        & \multicolumn{5}{c}{\multirow{2}{*}{--}}
        \\
        & $\gamma=$   & $p_{50th}$ & $p_{75th}$ & $p_{95th}$ 
        & \\
        \hline
        \multirow{2}{*}{$H_1$, $H_2$, $H_3$} 
        & $t_d=$      & $p_{50th}$ & $p_{75th}$ & $p_{95th}$
        & $x_{spec}=$ & 0.05       & 0.10       & 0.15 \\
        & $x_{min}=$  & 1.15       & 1.30       & 1.45
        \\
        \hline
    \end{tabular}
    \caption{Framework configurations consisting of proactive and reactive strategies, as defined in Section \ref{sec:methods}. $p_{xth}$ indicates value of the x-th percentile of the delays and operational area sizes of aeronautical users (see Appendix \hyperref[sec:users]{B} for more details).}
    \label{tab:configurations}
\end{table*}

In the second experiment, we introduce uncertainty and evaluate the proposed method when information about mobile users' trajectories or possible future locations is not accurately known when generating the baseline plan before operations, i.e., $|U_{info}|<|U_{tot}|$, as detailed in Section \ref{sec:setup}.



\begin{figure*}[t!]
    \centering
    \includegraphics[width=\textwidth]{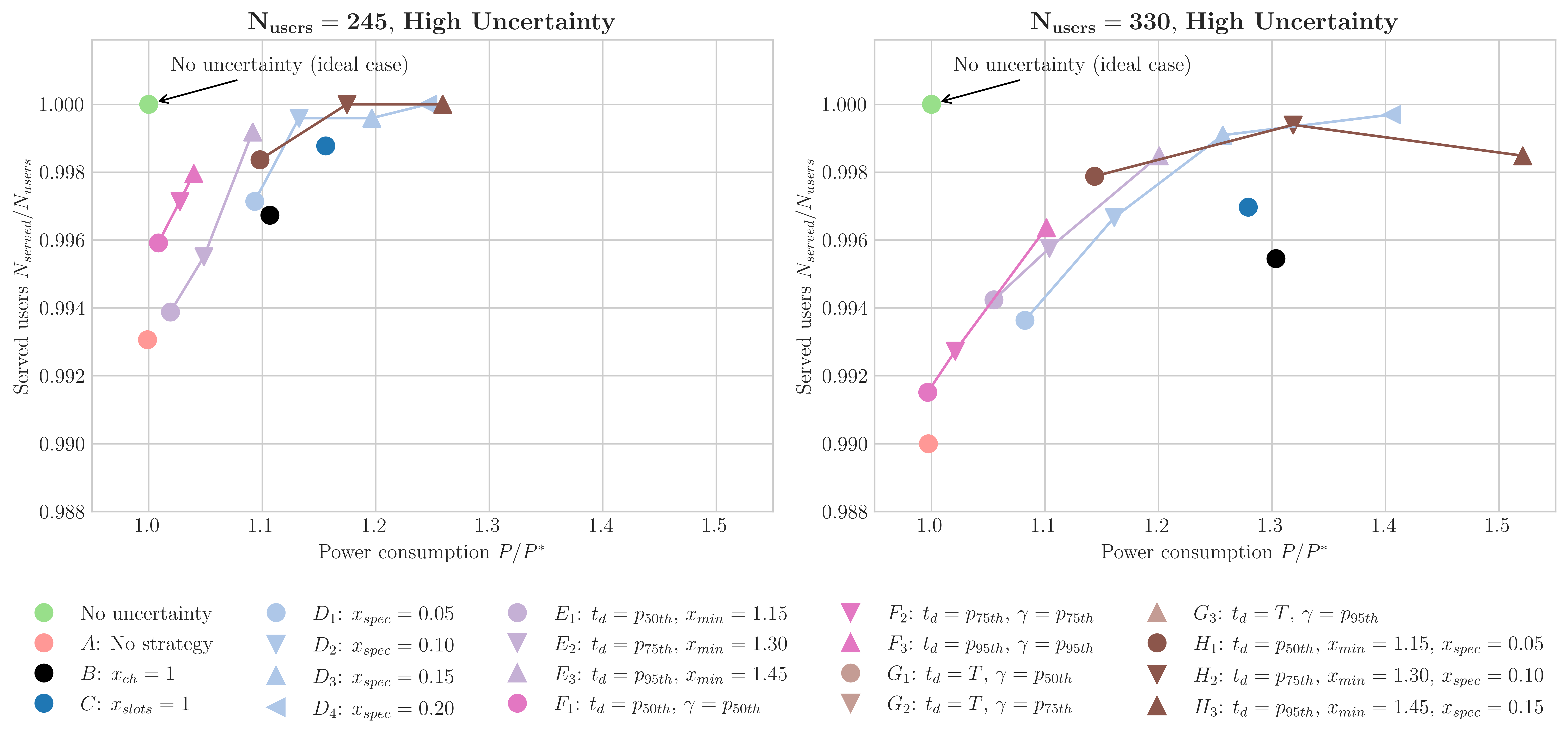}
    \caption{Average fraction of served users $N_{served}/N_{users}$ and normalized power consumption $P/P^*$ scenarios with 245 and 330 users and high uncertainty (results for 10 runs). $P^*$ is the power consumption of an ideal case with no uncertainty (in green).}
    \label{fig:r_tradeoffs}
\end{figure*}

\begin{table*}[t]
\footnotesize
    \centering
    \begin{tabular}{c|ccc|ccc}
\hline
\rowcolor{lightbg}
$N_{users}=330$
& \multicolumn{3}{c|}{\textbf{Low uncertainty}}
& \multicolumn{3}{c}{\textbf{High uncertainty}} \\
\hline
\rowcolor{lightbg}
& $N_{served}/N_{users}$ & $P/P^*$ & $N_{realloc}/N_{beams}$ 
& $N_{served}/N_{users}$ & $P/P^*$ & $N_{realloc}/N_{beams}$
\\
\rowcolor{lightbg}
\multirow{-2}{*}{Config.}
& Avg. (SD) & Avg. (SD) & Avg. (SD) & Avg. (SD) & Avg. (SD) & Avg. (SD) \\
\hline
$A$ & 0.996 (0.002) & \textbf{0.979} (0.050) & 0.097 (0.014) & 0.990 (0.005) & \textbf{0.997} (0.032) & 0.227 (0.017) \\
\hline
$B$ & 0.996 (0.004) & 1.340 (0.075) & 0.071 (0.009) & 0.995 (0.003) & 1.303 (0.117) & 0.169 (0.015) \\
\hline
$C$ & 0.998 (0.002) & 1.264 (0.091) & 0.068 (0.007) & 0.997 (0.004) & 1.279 (0.102) & 0.176 (0.009) \\
\hline
$D_1$ & 0.997 (0.002) & 1.064 (0.036) & 0.083 (0.009) & 0.994 (0.005) & 1.082 (0.055) & 0.202 (0.011) \\
$D_2$ & \textbf{1.000} (0.001) & 1.181 (0.075) & 0.084 (0.009) & 0.997 (0.003) & 1.161 (0.063) & 0.197 (0.012) \\
$D_3$ & \textbf{1.000} (0.000) & 1.301 (0.113) & 0.089 (0.011) & 0.999 (0.001) & 1.257 (0.071) & 0.203 (0.015) \\
$D_4$ & \textbf{1.000} (0.000) & 1.358 (0.097) & 0.097 (0.011) & \textbf{1.000} (0.001) & 1.405 (0.119) & 0.216 (0.009) \\
\hline
$E_1$ & 0.999 (0.001) & 1.069 (0.044) & 0.063 (0.006) & 0.994 (0.004) & 1.055 (0.056) & 0.187 (0.020) \\
$E_2$ & \textbf{1.000} (0.000) & 1.126 (0.051) & 0.051 (0.008) & 0.996 (0.004) & 1.104 (0.062) & 0.157 (0.013) \\
$E_3$ & 0.999 (0.001) & 1.214 (0.032) & \textbf{0.042} (0.006) & 0.998 (0.002) & 1.201 (0.075) & 0.097 (0.008) \\
\hline
$F_1$ & \textbf{1.000} (0.001) & 1.052 (0.152) & 0.065 (0.007) & 0.992 (0.005) & \textbf{0.997} (0.107) & 0.194 (0.013) \\
$F_2$ & \textbf{1.000} (0.000) & 1.047 (0.129) & 0.053 (0.006) & 0.993 (0.003) & 1.021 (0.099) & 0.155 (0.012) \\
$F_3$ & 0.999 (0.001) & 1.119 (0.170) & 0.048 (0.007) & 0.996 (0.003) & 1.101 (0.105) & 0.099 (0.010) \\
\hline
$G_1$ & 0.979 (0.006) & 1.615 (0.205) & 0.479 (0.029) & 0.984 (0.005) & 1.583 (0.286) & 0.487 (0.023) \\
$G_2$ & 0.980 (0.005) & 1.623 (0.209) & 0.487 (0.023) & 0.983 (0.005) & 1.574 (0.277) & 0.496 (0.022) \\
$G_3$ & 0.980 (0.004) & 1.614 (0.193) & 0.500 (0.025) & 0.983 (0.004) & 1.574 (0.271) & 0.509 (0.022) \\
\hline
$H_1$ & \textbf{1.000} (0.001) & 1.155 (0.073) & 0.053 (0.007) & 0.998 (0.002) & 1.144 (0.069) & 0.154 (0.010) \\
$H_2$ & \textbf{1.000} (0.000) & 1.350 (0.097) & 0.051 (0.005) & 0.999 (0.001) & 1.319 (0.107) & 0.126 (0.012) \\
$H_3$ & 0.999 (0.002) & 1.538 (0.135) & 0.045 (0.005) & 0.998 (0.004) & 1.521 (0.136) & \textbf{0.091} (0.010) \\
\hline
    \end{tabular}
    \caption{Results of the frequency assignment with uncertainty. $N_{users}$ is the number of users, $N_{served}/N_{users}$ is the fraction of successfully served users and $P/P^*$ is the power consumption normalized against the power required in an ideal case with no uncertainty $P^*$, and $N_{realloc}/N_{beams}$ is the number of reallocations in real-time normalized against the number of beams. The values in the table indicate the average and standard deviation (SD) obtained from 10 runs.}
    \label{tab:results_2}
\end{table*}

We perform 10 runs simulating different user distributions, and thus different user positions, service times, and unanticipated events, for each of the experiment configurations in Table \ref{tab:configurations}. This procedure is repeated for two uncertainty levels (low, high), and two different numbers of users (245 and 330) for a total of 760 runs. Figure \ref{fig:r_tradeoffs} shows the fraction of served users and the normalized power consumption for the different strategies under uncertainty ($U_{info}<U_{tot}$) . The numerical values can be found in Table \ref{tab:results_2}, where the number of reallocations is also included.

The results show that the best performance is obtained by an ideal case with no uncertainty (in green), which emphasizes the importance of generating a baseline plan with accurate user information. Generally, using proactive and reactive strategies improves the fraction of served users in all cases compared to config. $A$, which does not use any of the strategies S1-S6 to reduce uncertainty in the baseline plan or reserves resources.

Configurations $B$ and $C$, which reserve resources on a per-user basis, are outperformed in almost all cases due to their low reallocation success rate and spectrum utilization. In the scenario with 330 users and high uncertainty, less than 93.5\% of the reallocations are successful. Configurations $G_1$-$G_3$ offer the worst performance, as they include almost 4 times more constraints than necessary by not accurately taking into account the time that users will be active.
In the same scenario, reserving 20\% of the spectrum (config. $D_4$) achieves the highest fraction of served users (99.97\%) by using 40.5\% more power than an ideal case with no uncertainty. Reserving less spectrum (10\%) together with proactive strategies (config. $H_2$) achieves a similar fraction of served users (99.94\%) and power consumption but with 41.5\% fewer changes in real-time.

The results indicate that there is a tradeoff between the fraction of served users and the power consumption—for example, the same config. $D_4$ needs 41\% more power than config. $F_1$ to achieve a 0.8\% increase in served users. This difference tends to increase with dimensionality (see Figure \ref{fig:r_tradeoffs}) and slightly less with the level of uncertainty (see Table \ref{tab:results_2}), where the change is primarily in the fraction of served users (e.g., the fraction of served users for config. $F_1$ is reduced by 0.8\%).

While configurations with only proactive strategies (e.g., $E_2$, $F_2$) achieve good performance in cases with low uncertainty, the use of reactive strategies, more specifically reserving spectrum (configs. $D$, $H$), is crucial when uncertainty is high. This is partly due to the necessary flexibility, in terms of available resources, to serve users not included in the baseline plan.


\subsection{Discussion}

These results show the framework's capability to encode different strategies (S1-S6), which present different trade-offs in terms of the fraction of served users and power consumption (e.g., 41\% more power to serve 0.8\% more users), as well as the required number of reallocations.
It is important to highlight that the user base highly influences the strategies' performance. For instance, for users for which accurate information is known before operations (aeronautical users with possible trajectories), reactive strategies might not result in higher performance than having a more conservative allocation. In contrast, reserving resources can be a good solution in scenarios with users for which we do not know a priori when they will start service.

While in this work we assumed specific values to the strategies, the hyper-parameters present in the strategies can be adjusted by using user data or by simulation. For example, historical flight data can be analyzed for aeronautical users to characterize the delays and possible trajectories.
In contrast, when the hyper-parameters do not directly depend on the users (e.g., how much spectrum is reserved), configuring the strategies becomes more complex and requires performing experiments with different values.
Additionally, these strategies can be further configured by, for instance, changing the priority of users to access reserved resources.

Note that we have only tested two levels of dimensionality and uncertainty, in which we could meet all the demands assuming no uncertainty. For this reason, rather than characterize the framework's performance, the results should be used to understand what possible trade-offs the different strategies present, how they can change depending on the user distribution, and how to optimize them.

\section{Conclusions}
\label{sec:conclusions}



In this work, we have developed a frequency assignment framework to elaborate frequency plans valid for a specific period in the presence of mobile users. Our method consists of two stages. The proactive stage first generates a baseline plan before operations, and then the reactive stage re-optimizes it in real-time. The first experiment's results demonstrate that the framework can successfully capture mobility considerations when the information about the users is known a priori. The conclusions of the second experiment can be summarized as follows:
\begin{itemize}[itemsep=0em]
    \item Thanks to the combination of the proactive and reactive stages, the framework can successfully assign frequency resources for mobile users when not all information is known a priori, providing service to 99.97\% of the users in dense scenarios with over 900 beams.
    \item The framework's performance can be tuned by configuring different proactive and reactive strategies, presenting a trade-off between the fraction of served users, power consumption, and the number of changes in real-time. This trade-off is most significant in scenarios with more users and high uncertainty.
    \item Not taking into account the expected times that users will require service (e.g., expected flight departures) results in less efficient allocations.
    \item Reactive strategies consisting of reserving resources on a per-user basis are outperformed, whereas reserving part of the spectrum to reallocate users in real-time offers the highest fraction of served users.
\end{itemize}

Based on the results of this work, directions for future research include developing hybrid beam coverage solutions where fixed beams can serve mobile users, studying how mobile users can be efficiently routed to gateways, and extending the capabilities of the framework to operate under demand uncertainty.


\section*{Acknowledgments}
This work was supported by SES S.A.. The authors would like to thank SES S.A. for their input to this paper and their financial support. This work has also been partially supported by the mobility grants program of Centre de Formació Interdisciplinària Superior (CFIS) - Universitat Politècnica de Catalunya (UPC).

\section*{Appendix A. Power Calculation}
\label{sec:power}
The objective function in Eq. \ref{eq:problem} includes the power consumption of each beam. Since power equations are not linear \cite{Maral2010}, we precompute, for each of beam $i\in\{1,...,N_B\}$, the required power $P_i(f_i,b_i)$ for each possible frequency assignment.

We use the satellite communications models described in \cite{Paris2019}. For simplicity, we describe the method to compute the necessary power $P_i$ (in Watts) for a beam $i$ given its data rate demand $D_i$ (in bits/s) and a certain number of allocated frequency channels $b_i$. We assume the satellites use the MODCOD schemes defined in the standards DVB-S2 and DVB-S2X \cite{DigitalVideoBroadcastingDVB}. Given a roll-off factor $\alpha_i$ and channel width of $BW$ (in Hz), we compute the lower bound of the required spectral efficiency as
\begin{equation}
    \Gamma_{req} = \frac{D_i(1 + \alpha_i)}{b_i\cdot BW}
\end{equation}
We select the MODCOD with the lowest spectral efficiency such that $\Gamma \geq \Gamma_{req}$, from which the appropriate value for $E_b/N$ is obtained.

Since we have considered interference mitigation using the inter-group constraints, we assume interference is negligible, allowing us to compute the necessary $C/N_0$ as
\begin{equation}
    \left.\frac{C}{N_0}\right|_i = \left.\frac{E_b}{N}\right|_i \cdot \frac{D_i}{b_i\cdot BW}
\end{equation}
With $C/N_0$ (in dB), we can then compute the power as:
\begin{align}
    P_i = & \, \left.\frac{C}{N_0}\right|_i + OBO - G_{T_x} - G_{R_x}  \nonumber \\ & + \text{FSPL} + 10\log_{10}(kT_{sys})
\end{align}
where OBO is the power-amplifier output back-off, $G_{T_x}$ and $G_{R_x}$ are the transmitting and receiving antenna gains, respectively, $k$ is the Boltzmann constant, and $T_{sys}$ is the system temperature. FSPL and $L_{atm}$ account for the free-space path losses, respectively. Assuming that FSPL are significantly larger than atmospheric losses, and losses at the transmitting and receiving antennas, we neglect all the latter.

We compute a power value $P_i$ for each possible assignment of $b_i$, and repeat the process for all beams in the constellation.

\section*{Appendix B. User Distributions}
\label{sec:users}

\begin{figure}[t]
    \centering
    \begin{subfigure}[t]{0.49\linewidth}
        \centering
        \includegraphics[height=0.49\linewidth]{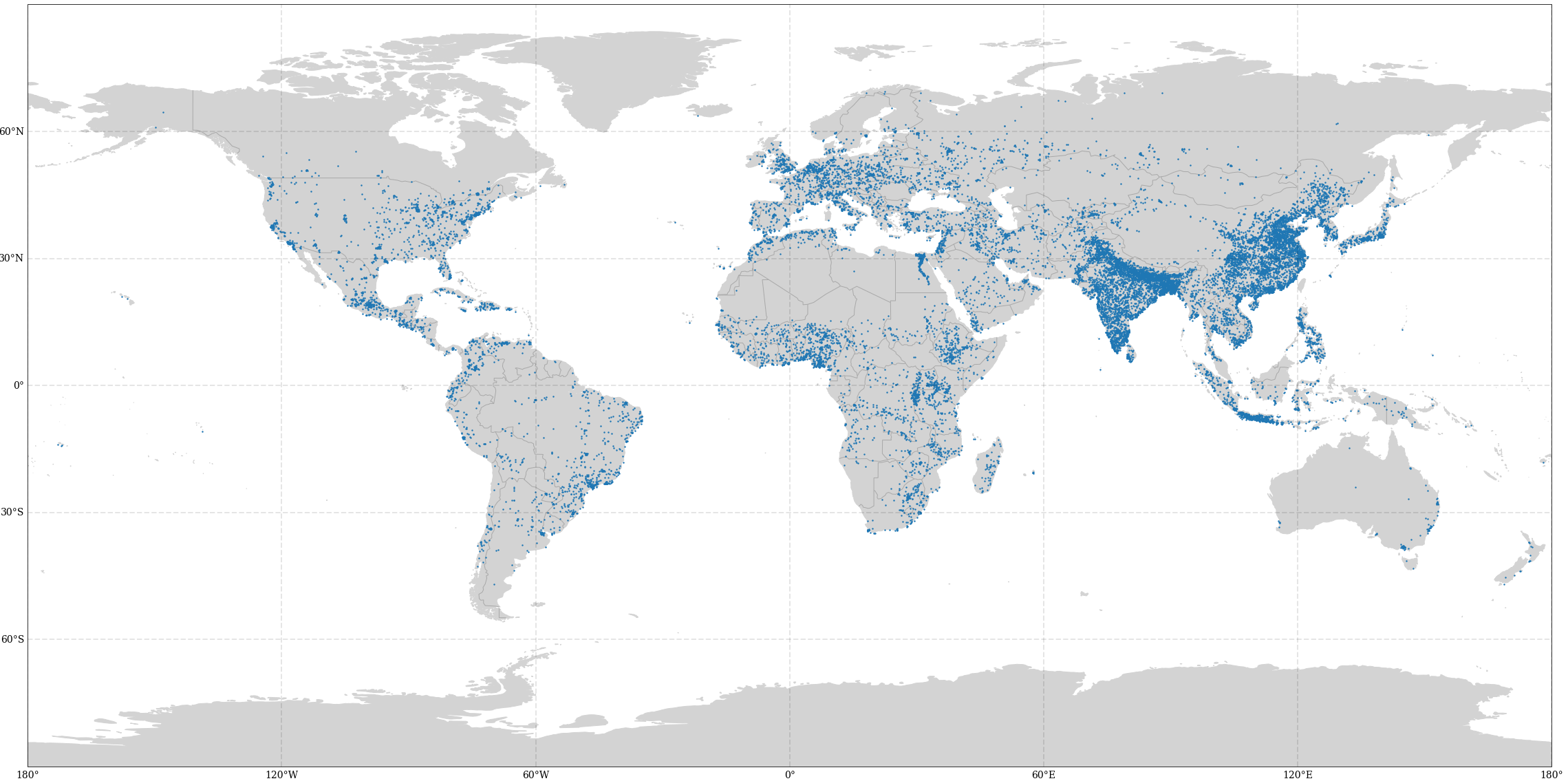}
        \caption{Fixed}
        \hfill
    \end{subfigure}
    \hfill
    \begin{subfigure}[t]{0.49\linewidth}
        \hfill
        \includegraphics[height=0.49\linewidth]{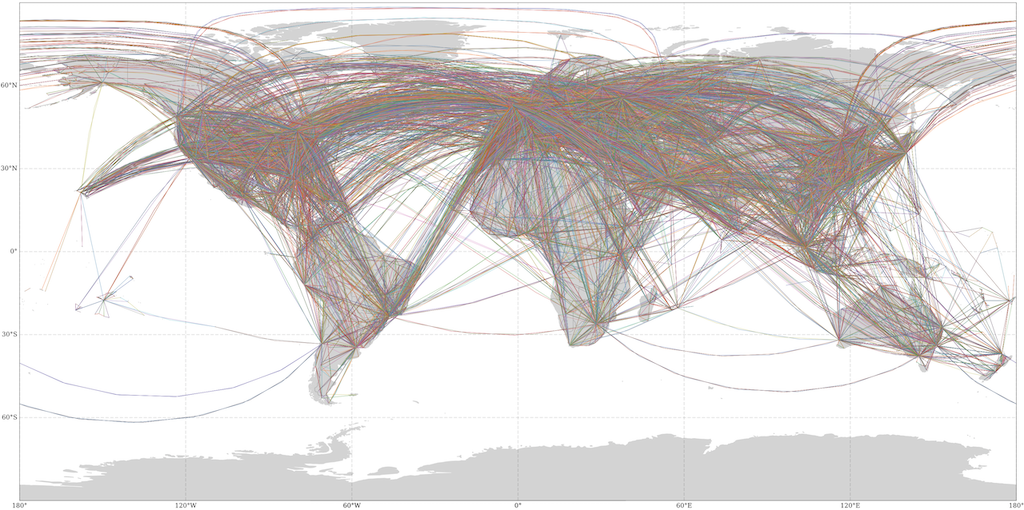}
        \caption{Aeronautical}
    \end{subfigure}
    \newline
    \centering
    \begin{subfigure}[t]{0.49\linewidth}
        \centering
        \includegraphics[height=0.49\linewidth]{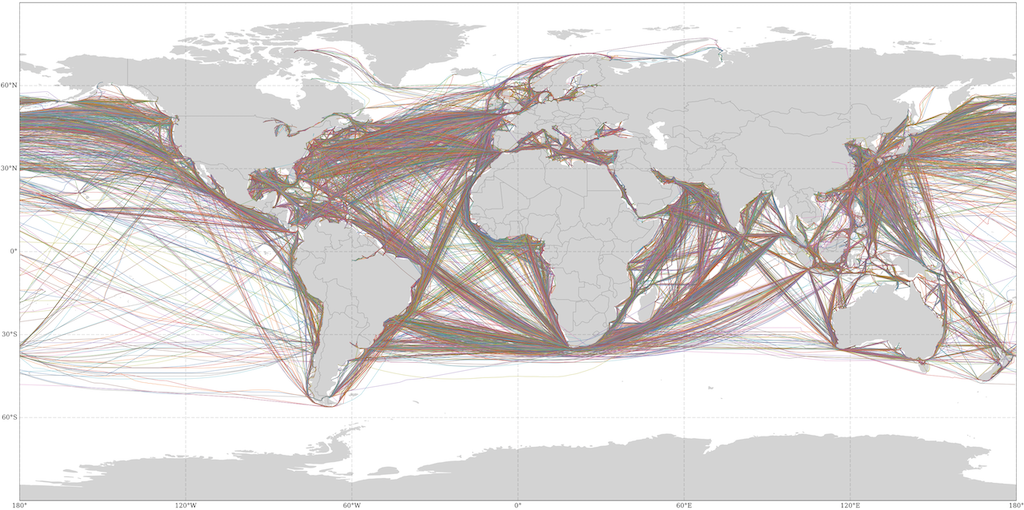}
        \caption{Maritime}
        \hfill
    \end{subfigure}
    \hfill
    \begin{subfigure}[t]{0.49\linewidth}
        \hfill
        \includegraphics[height=0.49\linewidth]{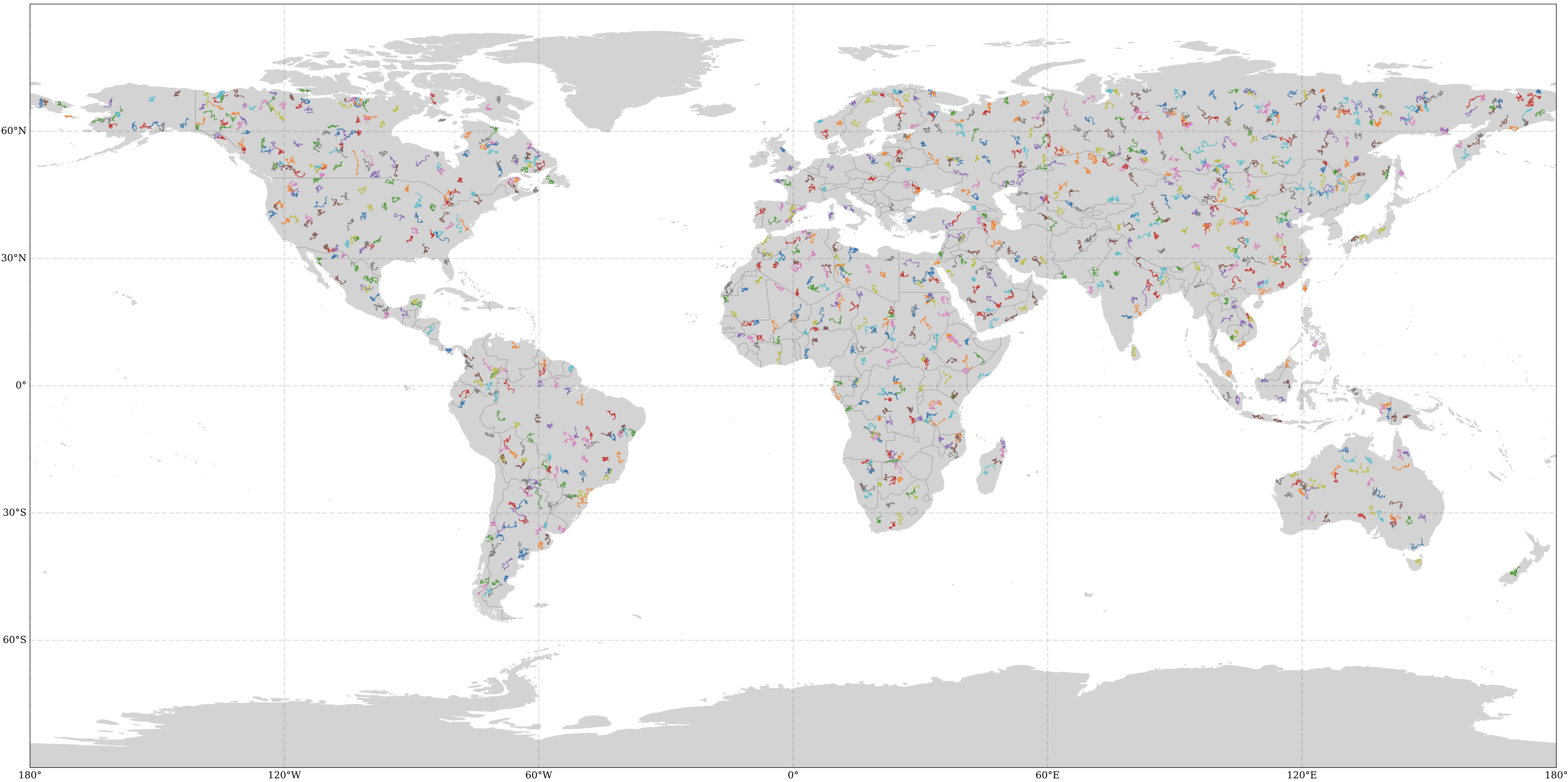}
        \caption{Land mobile}
    \end{subfigure}
    \caption{User datasets}
    \label{fig:user_datasets}
\end{figure}

In the experiments, the user positions are based on samples from datasets of population data \cite{CenterForInternationalEarthScienceInformationNetwork-CIESIN-ColumbiaUniversity2018}, flight routes and tracks \cite{JaniPatokallio,Eurocontrol2020}, and maritime routes \cite{Novikov2019}. Land mobile users are based on synthetic data. Table \ref{tab:number_users} shows the number of users by type, and Figure \ref{fig:user_datasets} shows the user distributions from which they are sampled.

\begin{table}[t]
    \footnotesize
    \centering
    \begin{tabular}{lcc}
        \hline
        \rowcolor{lightbg}
        User type    & Scenario 1 & Scenario 2 \\
        \hline
        Fixed        & 25         & 50         \\
        Aeronautical & 200        & 250        \\
        Maritime     & 10         & 15         \\
        Land mobile  & 10         & 15         \\
        \hline
        Total        & 245        & 330        \\
        \hline
    \end{tabular}
    \caption{Number of users by type.}
    \label{tab:number_users}
\end{table}

\begin{figure}[t]
    \centering
    \includegraphics[width=1\linewidth]{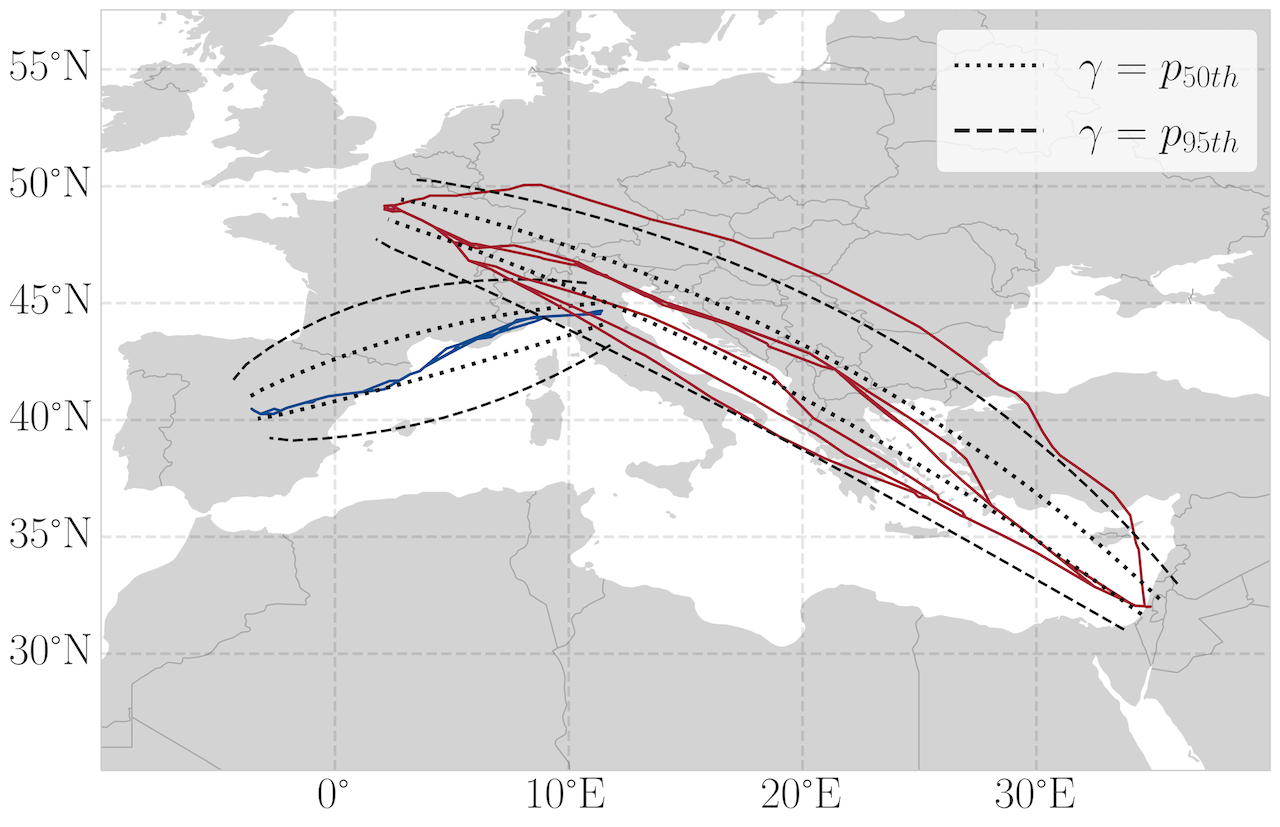}
    \caption{Multiple trajectories and possible operational areas (Section \ref{sec:methods}, strategy S3) for aeronautical users.}
    \label{fig:aeronautical}
\end{figure}

The uncertainty in the scenarios is based on:
\begin{itemize}[itemsep=0em]
    \item The \textit{fraction of land mobile users} in not included in $U_{info}$, being 25\% and 75\% for low and high uncertainty, respectively.
    \item The \textit{magnitude of the delay} with respect to the expected service time of \textit{aeronautical users}.
    \item The \textit{distance between possible trajectories} of \textit{aeronautical users}.
\end{itemize}
An example of the latter is shown in Figure \ref{fig:aeronautical}, where the blue flight would be considered to have low uncertainty in its trajectory, whereas the red flight would be classified with high uncertainty.

Figure \ref{fig:aeronautical} also depicts how operational areas $\mathcal{P}_i(t_i)$ (introduced in Section \ref{sec:methods}, strategy S3) are defined based on a parameter $\gamma$ when computing the constraints for aeronautical users in the proactive assignment stage (explained in Section \ref{sec:methods}). The percentiles $p_{xth}$ are computed based on the magnitude of the distance between past flight trajectories between two points. When no operational area is specified and information about only a user's start and end position is known, the constraints are computed assuming the shortest path between those points.

\bibliographystyle{ieeetr}
\bibliography{main}

\begin{thebibliography}{10}

\bibitem{NorthernSkyResearch2022}
{Northern Sky Research}, ``{{NSR}}'s {{Aero Satcom Report Sees Recovering IFC Market Generating}} \$48 {{Billion}} through {{Decade}},'' Apr. 2022.

\bibitem{NorthernSkyResearch2022b}
{Northern Sky Research}, ``{{NSR}}: {{Maritime Users}} to {{Increase Satellite Constellations Spending}} by 16x,'' July 2022.

\bibitem{Pachler2021a}
N.~Pachler, I.~{del Portillo}, E.~F. Crawley, and B.~G. Cameron, ``An {{Updated Comparison}} of {{Four Low Earth Orbit Satellite Constellation Systems}} to {{Provide Global Broadband}},'' in {\em 2021 {{IEEE International Conference}} on {{Communications Workshops}} ({{ICC Workshops}})}, pp.~1--7, June 2021.

\bibitem{NorthernSkyResearch2022c}
{Northern Sky Research}, ``{{NSR}}'s {{Satellite Capacity Report Sees Industry Moving Past COVID-19 Contraction}} to {{Drive}} \${{207B}} in {{Revenue Amidst Competition}}, {{Innovation}} and {{Risk-Taking}},'' July 2022.

\bibitem{InternationalTelecommunicationUnion2022}
{International Telecommunication Union}, ``Satellite issues: {{Earth}} stations in motion ({{ESIM}}).'' \url{https://www.itu.int:443/en/mediacentre/backgrounders/Pages/Earth-stations-in-motion-satellite-issues.aspx}, Mar. 2022.

\bibitem{SESS.A.}
{SES S.A.}, ``U.{{S}}. {{FCC Expands Market Access}} for {{SES O3b MEO Constellation}}.'' https://www.ses.com/press-release/us-fcc-expands-market-access-ses-o3b-meo-constellation.

\bibitem{SpaceExplorationHoldingsLLC}
{Space Exploration Holdings, LLC}, ``Application for {{Fixed Satellite Service}} by {{Space Exploration Holdings}}, {{LLC}} [{{SAT-LOA-20161115-00118}}].'' https://fcc.report/IBFS/SAT-LOA-20161115-00118.

\bibitem{Al-Hraishawi2021}
H.~{Al-Hraishawi}, H.~Chougrani, S.~Kisseleff, E.~Lagunas, and S.~Chatzinotas, ``A {{Survey}} on {{Non-Geostationary Satellite Systems}}: {{The Communication Perspective}},'' July 2021.

\bibitem{Markovitz2022}
O.~Markovitz and M.~Segal, ``{{LEO}} satellite beam management algorithms,'' {\em Computer Networks}, vol.~214, p.~109160, Sept. 2022.

\bibitem{Garau-Luis2021}
J.~J. {Garau-Luis}, S.~Eiskowitz, N.~Pachler, E.~Crawley, and B.~Cameron, ``Towards {{Autonomous Satellite Communications}}: {{An AI-based Framework}} to {{Address System-level Challenges}},'' Dec. 2021.

\bibitem{Guerster2019}
M.~Guerster, J.~Jose Garau~Luis, E.~Crawley, and B.~Cameron, ``Problem representation of dynamic resource allocation for flexible high throughput satellities,'' in {\em 2019 {{IEEE Aerospace Conference}}}, pp.~1--8, Mar. 2019.

\bibitem{Panthi2016}
S.~Panthi, {\em Dynamic {{Resource Management}} in {{Future Satellite Systems}} to {{Improve Resource Utilization}}}.
\newblock PhD thesis, UC Irvine, 2016.

\bibitem{McLain2013}
C.~McLain, S.~Panthi, and J.~King, ``Designing {{Satellites}} for the {{Broadband Aero Market}},'' in {\em 31st {{AIAA International Communications Satellite Systems Conference}}}, ({Florence, Italy}), {American Institute of Aeronautics and Astronautics}, Oct. 2013.

\bibitem{McLain2017}
C.~McLain and J.~King, ``Future {{Ku-Band Mobility Satellites}},'' in {\em 35th {{AIAA International Communications Satellite Systems Conference}}}, ({Trieste, Italy}), {American Institute of Aeronautics and Astronautics}, Oct. 2017.

\bibitem{Mizuike1989}
T.~Mizuike and Y.~Ito, ``Optimization of frequency assignment,'' {\em IEEE Transactions on Communications}, vol.~37, pp.~1031--1041, Oct. 1989.

\bibitem{Funabiki1997}
N.~Funabiki and S.~Nishikawa, ``A gradual neural-network approach for frequency assignment in satellite communication systems,'' {\em IEEE Transactions on Neural Networks}, vol.~8, pp.~1359--1370, Nov. 1997.

\bibitem{Wang2011}
J.~Wang, Y.~Cai, and J.~Yin, ``Multi-start stochastic competitive {{Hopfield}} neural network for frequency assignment problem in satellite communications,'' {\em Expert Systems with Applications}, vol.~38, pp.~131--145, Jan. 2011.

\bibitem{Salcedo-Sanz2005}
S.~{Salcedo-Sanz} and C.~{Bouso{\~n}o-Calz{\'o}n}, ``A {{Hybrid Neural-Genetic Algorithm}} for the {{Frequency Assignment Problem}} in {{Satellite Communications}},'' {\em Applied Intelligence}, vol.~22, pp.~207--217, May 2005.

\bibitem{Wang2015}
J.~Wang and Y.~Cai, ``Multiobjective evolutionary algorithm for frequency assignment problem in satellite communications,'' {\em Soft Computing}, vol.~19, pp.~1229--1253, May 2015.

\bibitem{Kiatmanaroj2012a}
K.~Kiatmanaroj, C.~Artigues, L.~Houssin, and F.~Messine, ``Frequency allocation in a {{SDMA}} satellite communication system with beam moving,'' in {\em 2012 {{IEEE International Conference}} on {{Communications}} ({{ICC}})}, pp.~3265--3269, June 2012.

\bibitem{Ortiz-Gomez2021}
F.~G. {Ortiz-Gomez}, D.~Tarchi, R.~Mart{\'i}nez, A.~{Vanelli-Coralli}, M.~A. {Salas-Natera}, and S.~{Landeros-Ayala}, ``Convolutional {{Neural Networks}} for {{Flexible Payload Management}} in {{VHTS Systems}},'' {\em IEEE Systems Journal}, vol.~15, pp.~4675--4686, Sept. 2021.

\bibitem{Hu2020}
X.~Hu, X.~Liao, Z.~Liu, S.~Liu, X.~Ding, M.~Helaoui, W.~Wang, and F.~M. Ghannouchi, ``Multi-{{Agent Deep Reinforcement Learning-Based Flexible Satellite Payload}} for {{Mobile Terminals}},'' {\em IEEE Transactions on Vehicular Technology}, vol.~69, pp.~9849--9865, June 2020.

\bibitem{Alberti2010}
X.~Alberti, J.~M. Cebrian, A.~Del~Bianco, Z.~Katona, J.~Lei, M.~A. {Vazquez-Castro}, A.~Zanus, L.~Gilbert, and N.~Alagha, ``System capacity optimization in time and frequency for multibeam multi-media satellite systems,'' in {\em 2010 5th {{Advanced Satellite Multimedia Systems Conference}} and the 11th {{Signal Processing}} for {{Space Communications Workshop}}}, pp.~226--233, Sept. 2010.

\bibitem{Cocco2018}
G.~Cocco, T.~De~Cola, M.~Angelone, Z.~Katona, and S.~Erl, ``Radio {{Resource Management Optimization}} of {{Flexible Satellite Payloads}} for {{DVB-S2 Systems}},'' {\em IEEE Transactions on Broadcasting}, vol.~64, pp.~266--280, June 2018.

\bibitem{Paris2019}
A.~Paris, I.~Del~Portillo, B.~Cameron, and E.~Crawley, ``A {{Genetic Algorithm}} for {{Joint Power}} and {{Bandwidth Allocation}} in {{Multibeam Satellite Systems}},'' in {\em 2019 {{IEEE Aerospace Conference}}}, ({Big Sky, MT, USA}), pp.~1--15, {IEEE}, Mar. 2019.

\bibitem{Abdu2021a}
T.~S. Abdu, S.~Kisseleff, E.~Lagunas, and S.~Chatzinotas, ``Flexible {{Resource Optimization}} for {{GEO Multibeam Satellite Communication System}},'' {\em IEEE Transactions on Wireless Communications}, pp.~1--1, June 2021.

\bibitem{Abe2018}
Y.~Abe, H.~Tsuji, A.~Miura, and S.~Adachi, ``Frequency {{Resource Allocation}} for {{Satellite Communications System Based}} on {{Model Predictive Control}} and {{Its Application}} to {{Frequency Bandwidth Allocation}} for {{Aircraft}},'' in {\em 2018 {{IEEE Conference}} on {{Control Technology}} and {{Applications}} ({{CCTA}})}, pp.~165--170, Aug. 2018.

\bibitem{Abdu2022}
T.~S. Abdu, S.~Kisseleff, E.~Lagunas, S.~Chatzinotas, and B.~Ottersten, ``Demand and {{Interference Aware Adaptive Resource Management}} for {{High Throughput GEO Satellite Systems}},'' {\em IEEE Open Journal of the Communications Society}, vol.~3, pp.~759--775, 2022.

\bibitem{Kisseleff2019}
S.~Kisseleff, B.~Shankar, D.~Spano, and J.-D. Gayrard, ``A new optimization tool for mega-constellation design and its application to trunking systems,'' in {\em Advances in {{Communications Satellite Systems}}. {{Proceedings}} of the 37th {{International Communications Satellite Systems Conference}} ({{ICSSC-2019}})}, pp.~1--15, Oct. 2019.

\bibitem{PachlerdelaOsa2021}
N.~{Pachler de la Osa}, M.~Guerster, I.~Portillo~Barrios, E.~Crawley, and B.~Cameron, ``Static beam placement and frequency plan algorithms for {{LEO}} constellations,'' {\em International Journal of Satellite Communications and Networking}, vol.~39, pp.~65--77, Jan. 2021.

\bibitem{Garau-Luis2022}
J.~J. {Garau-Luis}, S.~Aliaga, G.~Casadesus, N.~Pachler, E.~Crawley, and B.~Cameron, ``Frequency {{Plan Design}} for {{Multibeam Satellite Constellations Using Linear Programming}},'' Apr. 2022.

\bibitem{pachler22b}
N.~Pachler, {\em A Complete Resource Allocation Framework for Flexible High Throughput Satellite Constellations}.
\newblock PhD thesis, Massachusetts Institute of Technology Department of Aeronautics and Astronautics, {Cambridge, Massachusetts}, May 2022.

\bibitem{DelRe1993}
E.~Del~Re, R.~Fantacci, and G.~Giambene, ``Performance analysis of a dynamic channel allocation technique for terrestrial and satellite mobile cellular networks,'' in {\em Proceedings of {{GLOBECOM}} '93. {{IEEE Global Telecommunications Conference}}}, pp.~1698--1702 vol.3, Nov. 1993.

\bibitem{Maral1998}
G.~Maral, J.~Restrepo, E.~{del Re}, R.~Fantacci, and G.~Giambene, ``Performance analysis for a guaranteed handover service in an {{LEO}} constellation with a "satellite-fixed cell" system,'' {\em IEEE Transactions on Vehicular Technology}, vol.~47, pp.~1200--1214, Nov. 1998.

\bibitem{Zheng2020}
F.~Zheng, Z.~Pi, Z.~Zhou, and K.~Wang, ``{{LEO Satellite Channel Allocation Scheme Based}} on {{Reinforcement Learning}},'' {\em Mobile Information Systems}, vol.~2020, p.~e8868888, Dec. 2020.

\bibitem{Davari2019}
M.~Davari and E.~Demeulemeester, ``The proactive and reactive resource-constrained project scheduling problem,'' {\em Journal of Scheduling}, vol.~22, pp.~211--237, Apr. 2019.

\bibitem{Habibi2018}
F.~Habibi, F.~Barzinpour, and S.~J. Sadjadi, ``Resource-constrained project scheduling problem: Review of past and recent developments,'' {\em Journal of Project Management}, pp.~55--88, 2018.

\bibitem{FederalCommunicationsCommission2020}
{Federal Communications Commission}, ``Application for {{Fixed Satellite Service Other}} by {{SpaceX Services}}, {{Inc}}. [{{SES-LIC-INTR2021-00934}}].'' https://fcc.report/IBFS/SES-LIC-INTR2021-00934, May 2020.

\bibitem{Maral2010}
G.~Maral, M.~Bousquet, and Z.~Sun, {\em Satellite {{Communications Systems}}: {{Systems}}, {{Techniques}} and {{Technology}}}.
\newblock {Chichester, West Sussex, U.K}: {Wiley}, 5th edition~ed., Feb. 2010.

\bibitem{DigitalVideoBroadcastingDVB}
{Digital Video Broadcasting (DVB)}, ``Second generation framing structure, channel coding and modulation systems for {{Broadcasting}}, {{Interactive Services}}, {{News Gathering}} and other broadband satellite applications; {{Part}} 2: {{DVB-S2 Extensions}} ({{DVB-S2X}}).'' https://dvb.org/?standard=second-generation-framing-structure-channel-coding-and-modulation-systems-for-broadcasting-interactive-services-news-gathering-and-other-broadband-satellite-applications-part-2-dvb-s2-extensions.

\bibitem{CenterForInternationalEarthScienceInformationNetwork-CIESIN-ColumbiaUniversity2018}
{Center For International Earth Science Information Network-CIESIN-Columbia University}, ``Gridded {{Population}} of the {{World}}, {{Version}} 4 ({{GPWv4}}): {{Population Count}}, {{Revision}} 11,'' 2018.

\bibitem{JaniPatokallio}
{Jani Patokallio}, ``{{OpenFlights}}: {{Airport}} and airline data.'' https://openflights.org/data.html.

\bibitem{Eurocontrol2020}
{Eurocontrol}, ``R\&{{D}} data archive.'' https://www.eurocontrol.int/dashboard/rnd-data-archive, Sept. 2020.

\bibitem{Novikov2019}
A.~Novikov, ``Creating sea routes from the sea of {{AIS}} data..'' https://towardsdatascience.com/creating-sea-routes-from-the-sea-of-ais-data-30bc68d8530e, June 2019.

\end{thebibliography}

\end{document}